\documentclass[twocolumn]{emulateapj}
\usepackage{aas_macros}
\usepackage{graphicx}
\usepackage{epstopdf}
\newcommand{\be}{\begin{equation}}
\newcommand{\ee}{\end{equation} }
\newcommand{\ba}{\begin{eqnarray}}
\newcommand{\ea}{\end{eqnarray}}

\newcommand{\nn}{\mbox{} \nonumber \\ \mbox{} }
\newcommand{\kB}{k_{\rm B}}

\begin{document}
\shorttitle{Pulse Structure of Hot Electromagnetic Outflows}
\shortauthors{Thompson \& Gill}
\title{Pulse Structure of Hot Electromagnetic Outflows with Embedded Baryons}
\author{Christopher Thompson and Ramandeep Gill}
\affil{Canadian Institute for Theoretical Astrophysics, 60 St. 
George St., Toronto, ON M5S 3H8, Canada}
\begin{abstract}
Gamma-ray bursts (GRBs) show a dramatic pulse structure that requires
bulk relativistic motion, but whose physical origin has remained murky.
We focus on a hot, magnetized jet that is emitted by a black hole and interacts
with a confining medium.  Strongly relativistic expansion of the magnetic field,
as limited by a corrugation instability, may commence only after it forms a thin shell.
Then the observed $T_{90}$ burst duration is dominated by the
curvature delay, and null periods arise from angular inhomogeneities,
not the duty cycle of the engine.  We associate the $O(1)$ s timescale observed 
in the pulse width distribution of long GRBs with the
collapse of the central 2.5-3$M_\odot$ of a massive stellar core.  
A fraction of the baryons are shown to be embedded in
the magnetized outflow by the hyper-Eddington radiation flux; they strongly 
disturb the magnetic field after the compactness drops below 
$\sim 4\times 10^3(Y_e/0.5)^{-1}$.  The high-energy photons so created have a compressed 
pulse structure.  Delayed breakout of magnetic field from heavier baryon shells is 
also a promising approach to X-ray flares.   In the
second part of the paper, we calculate the imprint of an expanding, scattering 
photosphere on pulse evolution.  Two models for generating the high-energy spectral
tail are contrasted:  i) pair breakdown due to reheating of an optically thin 
pair plasma embedded in a thermal radiation field; and ii) continuous heating extending
from large to small scattering depth.  The second model is strongly inconsistent with the
observed hard-to-soft evolution in GRB pulses.  The first shows some quantitative differences
if the emission is purely spherical, but we show that finite shell width, mild departures from spherical
curvature, and latitudinal Lorentz factor gradients have interesting effects.
\end{abstract}
\keywords{MHD --- plasmas --- radiative transfer --- scattering --- gamma rays: bursts}

\maketitle
\section{Introduction}

The fast variability that is seen in gamma-ray bursts can only be produced by a 
relativistic outflow \citep{fenimore96,sari97}.   The evolution of individual gamma-ray pulses
provides some evidence for relativistically aberrated emission from a curved shell, although
not all details agree with the simplest model of optically thin emission from a spherically
symmetric outflow (e.g. \citealt{shenoy13}, and references therein).

In this paper, we consider two aspects of this variability:  angular variations in the
outflow, and the effects of photospheric scattering.  Our investigation is anchored by
focusing on hot, magnetized outflows that interact with ambient baryons \citep{thompson06}.  


The influences of radial and non-radial inhomogeneities on the gamma-ray lightcurve
are difficult to untangle, because relativistic beaming restricts our sampling of the emitting material.  
It has been noted that strong angular inhomogeneities could contribute to the wide luminosity
distribution of GRBs \citep{kumar_piran00}; and suggested that fast variability could result from
interaction with an external medium if the ejecta are strongly clumped on small scales \citep{heinz99}.  
Much attention has been given to radial variations in
kinetic energy flux \citep{rm94,kobayashi97,daigne98}, or radial striping of a magnetic field \citep{thompson94,spruit01,
zhang11,mckinney12}.

Concrete examples of outflows with both radial and non-radial structure have
been harder to construct.  The example studied here naturally has this property:
pulse formation is tied to the localized breakout of a hot and relativistically
magnetized fluid from a much denser matter shell.  

We find that the formation of distinct pulses does not depend on strong 
inhomogeneities such as local bursts of bulk relativistic motion driven by 
magnetic reconnection \citep{lyutikov03,narayan09,lazar09}, or inverse 
Compton emission beamed along the local magnetic field direction 
\citep{thompson06}.  But the fastest (e.g. sub-pulse) variability is
certainly enhanced by such mechanisms, and they may be required in the
most strongly variable GRBs.

In this paper we consider a few related questions:

1. How is the pulse structure of a GRB influenced by the inevitable two-component composition
of the outflow, resulting from the entrainment of baryons from a confining medium?   Dissipation
is driven by this interaction at the initial breakout of the relativistic material
(\citealt{tmr07,lmmb13,levinson13,tg13}, hereafter Paper I, \citealt{eichler14}).  A second phase of dissipation is
concentrated at a much larger radius, where
the photon compactness drops below a critical level, leading to strong differential motions 
between the baryons and a magnetized component (\citealt{thompson06,gt14}, hereafter Paper II). 

Here we argue that in many GRBs, 
the observed $T_{90}$ burst duration does not represent the active period of the central engine.  
Several lines of evidence, especially null periods in GRB light curves and delayed flaring activity
in long and short GRBs, are consistent with the delayed breakout of magnetized material from a confining baryonic shell.
This non-spherical structure of the flow has an obvious potential impact on the evolution of gamma-ray pulses, including
on detailed diagnostics of shell curvature.  

In long GRBs, a physical connection can be drawn between 
the collapse time of the material that accretes onto a black hole, and the $O(1)$ s timescale that is obtained from the
pulse width distribution \citep{norris96} and the temporal power spectrum \citep{bss00}.   Consideration of the propagation
time of jets through CO and He cores shows that strong focusing is required, with isotropic-equivalent  $L_{\rm iso}
\gtrsim 10^{53}$ erg s$^{-1}$, corresponding to opening angle $\theta_{\rm j} \lesssim 0.1$ rad.  We infer that
most long GRBs are powered by the magnetized cocoon, not by directly escaping jet material.

2. What are the limitations on fast variability in an outflow that begins
free expansion only at a relatively large radius \citep{eichler00,thompson06,russo13a,russo13b}?  We find that
the high-energy
photons that are formed during a delayed pair breakown (Paper II) can vary on 
a modest fraction of a pulse width, as determined by the engine activity and magnetic field breakout.  Magnetic reconnection tends to 
freeze out in this second stage and so may play only a limited role in forming the non-thermal high-energy 
spectral tail.  

3. What is the imprint of an offset scattering photosphere on pulse evolution?  The output spectrum
resulting from distributed heating around the scattering photosphere has been calculated
in various contexts:  in a baryonic plasma of variable scattering depth but in the static approximation \citep{peer06}; 
in an expanding baryonic plasma starting inside the photosphere and continuing to a low scattering 
depth \citep{giannios06b,giannios08,asano13}; in a strongly magnetized pair plasma of a high compactness but low effective 
temperature (Paper I); and in a plasma of low initial scattering depth but moderately high compactness, 
leading to pair breakdown (Paper II).  These last two calculations represent, first, the formation of the spectral peak
and low-energy spectral tail; and, second, the high-energy spectral tail.  

A principal goal here is to compare the pulse behavior resulting from continuous heating across a photosphere, with
that resulting from rapid pair breakdown in an optically thin flow.  We show that the first model is strongly 
inconsistent with the observed pulse behavior.  Delayed pair breakdown, if concentrated in radius, produces the 
observed hard-to-soft evolution and decrease of peak energy with time.  But we find that residual scattering at a photosphere
does tend to steepen pulse evolution.  Other effects, including non-spherical shell curvature and latitudinal
Lorentz factor gradients, act in the opposite direction and may be required even in the case of optically thin emission.

\subsection{Plan  of the Paper}

We frame our approach to pulse formation in GRBs in Section \ref{s:clues}
by dividing the problem into separate puzzles, and connecting them with the dynamical behavior
of collapsing stellar cores and hot electromagnetic outflows.  Then in Section \ref{s:baryon} 
we consider the entrainment of baryons at the jet head, due to a combination of 
corrugation instability and radiation pressure.  
Pulse variability produced
by the breakout of a magnetized fluid from a baryon shell is addressed in 
Section \ref{s:vary}, where we consider both the effect of a geometrically
thin breakout shell, and the faster variability that may result from delayed
reheating after the outflow has accelerated to a Lorentz factor $\sim 10^2-10^3$.
We extend our considerations of GRB prompt emission to delayed X-ray flares in Section \ref{s:afterglow}.  
The imprint of photospheric scattering, and delayed pair breakdown, on pulse
evolution is explored with Monte Carlo calculations in Section \ref{s:shell}, and
the result of pair breakdown at low optical depth is contrasted with corona-like models.
Section \ref{s:discuss} summarizes our results and makes note of some outstanding problems.
The Appendix details the Monte Carlo code.

\section{Challenges for Understanding \\ Gamma-ray Burst Variability}\label{s:clues}

We begin by considering a few outstanding observational issues,
along with their theoretical implications.

\subsection{Origin of Null Periods in Long GRBs}\label{s:null}

In many GRBs one observes
long intervals between pulses when the flux in the gamma-ray band appears to be dominated by off-axis emission.  Sometimes the
pulses are far enough separated that there is no measureable flux between them (e.g. \citealt{nakar02},
and references therein).

\subsubsection{Inconsistency of a Radially Modulated Outflow}

Such null intervals present a major challenge to any model of GRB variability that invokes {\it radial}
structure in a relativistic outflow.  An inconsistency could easily be avoided if the engine were `unclothed'
and fed by an intermittent accretion flow.  The difficulty arises from the interaction of the outflow
with a dense, confining medium:  either a stellar envelope \citep{paczynski98,macfadyen99}, or a neutron-rich
outflow from a merged binary neutron star \citep{dessart09}.  This medium is dense enough that the 
shocked jet creates a high-pressure cocoon (e.g. \citealt{matzner03,bromberg14}) that would fill in
any null periods in jet output.  This difficulty applies to any shock-based model, as well as one in
which radially separated slabs are strongly magnetized (e.g. \citealt{zhang11}).

\begin{figure}
\includegraphics[width=0.5\textwidth]{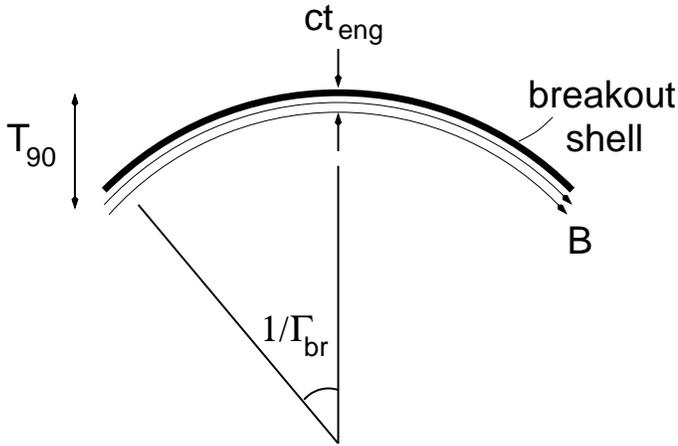}
\caption{A shell of relativistic magnetofluid is trapped behind a thin layer of baryonic material
that it has collected during breakout from a confining medium.  Both baryons and magnetofluid move at a similar
Lorentz factor $\sim \Gamma_{\rm br}$.  The magnetofluid was transported outward by a jet from
a black hole engine that was active for a time $t_{\rm eng}$.  Pinning by the baryons temporarily confines 
the magnetofluid into a layer of thickness $ct_{\rm eng}$.   Here the shell has expanded past the transition
to a `pancake' geometry, corresponding to a radius $> 2\Gamma_{\rm br}^2 ct_{\rm eng}$.  A delayed
corrugation mode, investigated in Section \ref{s:baryon}, then allows localized plumes of magnetofluid to
escape the baryon shell and expand to a much higher Lorentz factor.  These correspond to gamma-ray pulses
of duration $\sim t_{\rm eng}$.  The observer detects pulses over a wider time interval $T_{90}$, which 
is governed by the curvature delay, not by the outflow duration.}
\vskip .2in
\label{fig:shell}
\end{figure}

Consider a jet of angular radius $\theta_{\rm j}$ with a null interval $\Delta t \sim 1$ s in the relativistic flow.  
The pressurized cocoon fills in the jet if $\theta_{\rm j} r/c \lesssim \Delta t$, corresponding to a distance
$r \lesssim 1\times 10^{11}\,(\Delta t/1~{\rm s})(\theta_{\rm j}/0.1)^{-1}$ cm from the engine.  A cocoon structure must develop
well inside this radius.

If the cocoon material derived partly from the confining medium, or from a neutron-rich wind emitted by an 
orbiting torus, then it would only be mildly relativistic.  Outside breakout, it would interfere with the expansion of the 
relativistic jet fluid, and could easily suppress gamma-ray emission.  The radial layers would
come into causal contact outside a radius $\sim c\Delta t$, and beyond that point the combined inertia 
of the two-layer jet would approach that of the heavier component.  

Null intervals in the jet could, alternatively, be filled in only with shocked jet fluid and avoid
significant baryon contamination.  But then at least a thermal X-ray radiation field would be detectable in 
between isolated non-thermal pulses, carrying an energy flux {\it comparable} to that of the intervening gamma-ray pulses.

This argument also highlights a basic incompatibility between generating hard gamma-ray pulses with radial flow structure, and an
inverse-Compton origin of the gamma-rays.  Reheating of embedded $e^\pm$ could be concentrated in small radial steps 
after the outflow reaches high $\Gamma$; but in this situation, the soft photon field would have a smoother radial 
profile, being generated at lower $\Gamma$.  A nearly thermal radiation field would re-emerge in between the non-thermal pulses, carrying a comparable energy flux.
Broad-band null periods would be absent.

\subsubsection{Breakout of Magnetic Field from a \\ Thin, Curved Shell}

In this paper we investigate a magnetized outflow with significant {\it non-radial structure}.
After the outflow escapes the confining medium, it can still entrain enough baryons to prevent
further radial acceleration.  The flow evolves into a thin, magnetized shell of a radial
thickness $\Delta r \sim ct_{\rm eng}$, where $t_{\rm eng}$ is active period of the engine.  Such
a structure forms most easily if a magnetized cocoon is the source of the relativistic material
(see Section \ref{s:magcocoon}).  

The magnetic field is stretched in the non-radial direction as the shell expands.  The total electromagnetic energy contained at $\theta \leq \theta_j$ is
\be\label{eq:emag}
E_P \sim \pi\theta_{\rm j}^2 r^2 {B_\phi^2\over 4\pi} \Delta r \quad\quad (\Gamma \gg 1).
\ee
A key point is that this energy changes slowly with radius, insofar as the 
shell maintains nearly constant $\Delta r$.  That is the case if the the kinetic energy of the entrained baryons 
dominates the magnetic energy:  then the Alfv\'en speed 
$B_\phi'/(4\pi\rho')^{1/2} \ll c$ in the comoving frame.

Eventually the baryons swept up at the head of the shell develop a corrugation instability, 
but only beyond a radius $2\Gamma^2_{\rm br}\Delta r$.  The corresponding curvature delay associated
with isolated breakout of magnetofluid, at angles $\theta \lesssim 1/\Gamma$ from the observer-engine
axis, is
\be
t_{\rm curve} \sim {1\over 2}\theta^2 {R_{\rm br}\over c}  \sim {R_{\rm br}\over 2\Gamma^2_{\rm br}\Delta r} t_{\rm eng} > t_{\rm eng},
\ee
as depicted in Figure \ref{fig:shell}.  Throughout this paper, 
the subscript `br' labels breakout.  Details of the corrugation instability are examined in Section \ref{s:baryon}.

In extreme cases, enough baryonic material overlaps with the magnetic field that breakout of the magnetic field
is delayed beyond the point where thermal photons have decoupled and flow ahead of the shell.  Then 
breakout appears as pulsed emission during the X-ray afterglow, as we investigate in Section \ref{s:afterglow}.
An intermediate case has been considered by \cite{lyutikov03}, in which the breakout that powers the prompt gamma-ray
emission is delayed to a radius $\sim 10^{16}$ cm --- far enough out that strongly anisotropic emission in the comoving
frame must be invoked to explain variability on $\lesssim 1$ s timescales.  
Strongly delayed breakout is more
naturally associated with heavier baryon loadings and slower phenomena such as X-ray flashes.

\subsubsection{Comparison with Patchy Jet Model}

\cite{kumar_piran00} describe a kinematic `patchy shell' model which combines radial and angular structure.
A comparison of this model with the present approach is instructive.
The main goal in \cite{kumar_piran00} was to explain the broad luminosity distribution of GRBs, rather than the pulse structure,
which was incorporated in the standard way through a choice of radial shell structure.  In this model, bursts of gamma-ray 
emission result from collisions between discrete blobs of a size $\sim \Gamma^{-1}$, distributed randomly in 
broader spherical shells.  The shells are porous, so that two shells collide only over a fraction of their solid 
angle.  Collisions between small blobs with high $\Gamma$ (and, hence, high-luminosity GRBs) are rare due to
the small solid angle assumed for such blobs.   The active period of the engine is 
comparable to the observed $T_{90}$ gamma-ray duration, and the flow direction is uniformly radial.

In the present approach, the detection of gamma rays from a part of the outflow is determined by the flow geometry
at radius $R_{\rm br}$, not by chance events within the reheating zone.  Significant angular gradients in flow direction can
be maintained when the relativistic component begins its (nearly) free expansion at $R_{\rm br} > 2\Gamma_{\rm br}^2 ct_{\rm eng}$, seeded by a local instability.  

\subsection{Origin of $O(1)$ s Timescale in Long GRBs:  Pulse Duration and Power Spectral Break}\label{s:tbreak}

The power spectrum of long GRBs shows a distinct break at $\sim 1$ s
\citep{bss00}. Isolated pulses in bright, long GRBs have a distribution
of widths that peaks at a similar timescale \citep{norris96}.  
Furthermore, null periods and pulses show similar distributions
(e.g. \citealt{nakar02}, and references therein), which points to a characteristic physical scale in the outflow.
 
\begin{figure}
\includegraphics[width=0.5\textwidth]{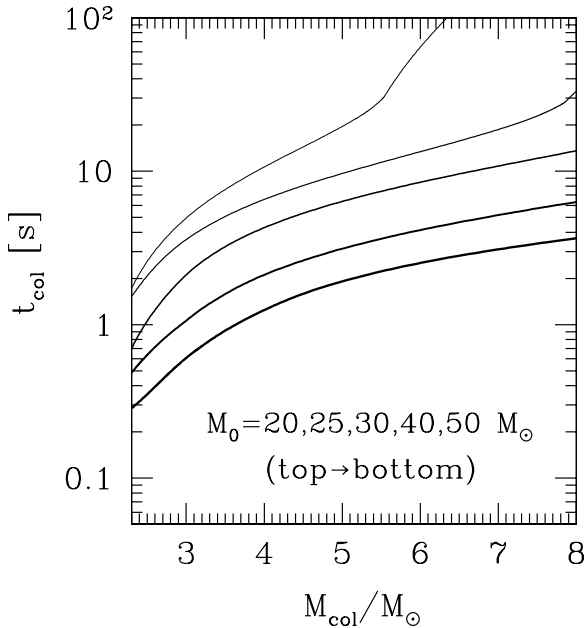}
\caption{Collapse time, defined as twice the free-fall time, equation (\ref{eq:tcol}), in the pre-collapse 
cores of massive stars.  Zero-age main sequence mass $M_0$, metallicity $10^{-3}$, models evolved by the MESA code
\citep{paxton13}.}
\vskip .2in
\label{fig:tcol}
\end{figure}

A similar timescale arises from a completely independent consideration.  The collapse time of the central core
of a massive star, containing $\gtrsim 2.5$-$3\,M_\odot$ and with enough material to form a black
hole, is typically less than the $\sim 3$-$30$ s duration of a long GRB.
For example, one finds $t_{\rm col} \sim 0.5$-$3$ s in cores forming within stars of zero-age main 
sequence mass $20$-$30\,M_\odot$ (Figure \ref{fig:tcol}).

\begin{figure}
\includegraphics[width=0.5\textwidth]{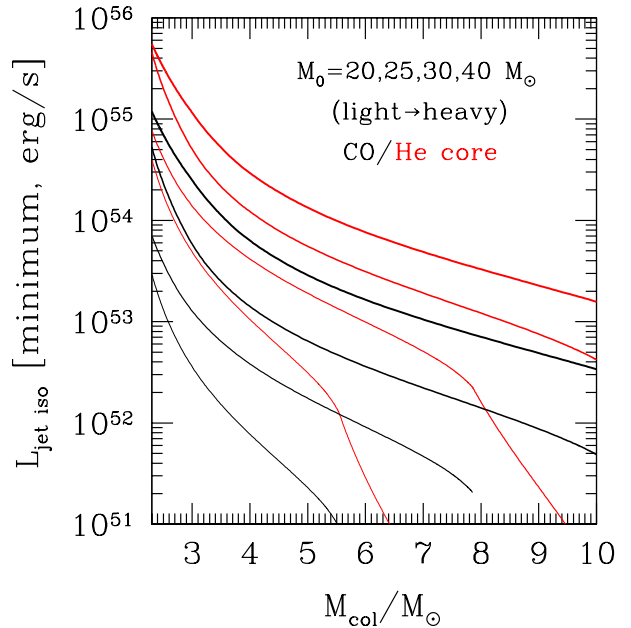}
\caption{Minimum isotropic-equivalent luminosity of a jet that punctures the CO core (black lines) or He core (red lines),
as a function of the central mass $M_{\rm col}$ that collapses during the active period of the jet (equation
(\ref{eq:lmin})).  Black curves end when
$M_{\rm col} > M_{\rm CO}$, the total CO mass at the time of core collapse.   Progenitor stars have metallicity 
$Z = 10^{-3}$ and zero-age main masses as labelled.}
\label{fig:ljet1}
\end{figure}
\begin{figure}
\includegraphics[width=0.5\textwidth]{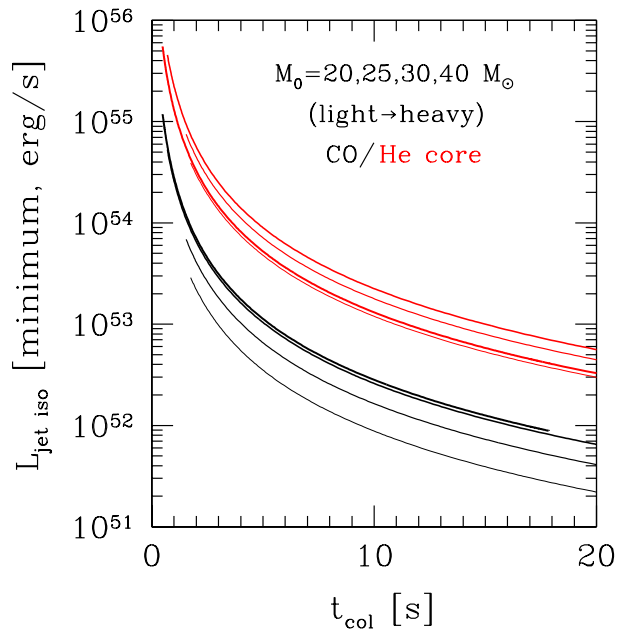}
\caption{Same as Figure \ref{fig:ljet1}, but now plotted in terms of the minimum active period of the jet $t_{\rm head}$,
equation (\ref{eq:thead}), set equal to the collapse time of the accreted mass.}
\label{fig:ljet2}
\vskip .2in
\end{figure}

One interpretation of these two facts is that accretion onto the black hole is cut off in a chaotic manner.
For example, an outflow from the neutron torus could interfere with accretion.  The characteristic timescale
for shutting off accretion would be the infall time of the uncollapsed material.  However there is
a weak motivation here for multiple cycles of outflow and accretion, given the much greater specific 
energy of the collapsed mass.  

A second interpretation, which we favor, is that the outflow duration at breakout
is significantly {\it shorter} than $T_{90}$, and better represented by the width of an individual
gamma-ray pulse.  

\subsection{Origin of GRB Luminosity Function: \\  Lateral Spreading of Focused Jets During Breakout}\label{s:magcocoon}

Long gamma-ray bursts have a very wide range of isotropic-equivalent energies
($E_{\rm \gamma,iso} \sim 10^{51}$-$10^{54}$ erg), especially when compared with the narrow range of binding energies of 
CO cores in evolved massive stars ($E_{\rm bind} \sim 1-4\times 10^{51}$ erg).  The significance
of this dichotomy has been obscured by the very large specific energy that is released by
accretion onto a black hole:  in principle, given the build-up of a sufficiently strong magnetic
field, a collimated jet could transmit a net energy $\gg E_{\rm bind}$.  Such
a hyper-energetic jet would, nonetheless, be surrounded by a broader outflow emitted by the surrounding 
neutron-rich torus (e.g. \citealt{sadowski13}), which in turn would interact with a collapsing stellar core.  

Therefore a feedback mechanism is present which limits the angle-integrated jet energy that
can be emitted by a collapsar \citep{tmr07}.   
Consider, for example, a jet of Lorentz factor $\Gamma$ and opening angle $\theta_{\rm j}$ in approximate transverse 
pressure balance with a trans-relativistic wind of pressure $P_{\rm ex}$.  The jet has a comoving pressure $\sim 
P_{\rm ex}$ and an enthalpy flux 
$\sim (2-4)\Gamma^2 P_{\rm ex}c$.  Its output, relative to the much broader wind, is
\be
{L_{\rm j}\over L_{\rm wind}} \sim (\Gamma\theta_{\rm j})^2.
\ee
Transverse pressure balance requires $\Gamma\theta_{\rm j} \lesssim 1$.  A broad wind therefore releases
comparable energy to the jet, and by inhibiting accretion establishes a connection between jet energy
and core binding energy.

Very high $E_{\rm \gamma,iso}$ can, of course, result from strong collimation.  Jet propagation
in an envelope is complicated by details of radial acceleration, as well as the interaction with
a cocoon of shocked jet material \citep{matzner03,bromberg14}.  It is somewhat easier to ask what
isotropic-equivalent luminosity a jet must attain to propagate through to the surface of a CO
core in the time that the core material collapses.  The head of a relativistic jet pushing through
dense stellar material moves radially at a speed
\be
v_{\rm head}(r) \simeq \left[{L_{\rm j,iso}\over 4\pi\rho_{\rm env}(r) r^2 c}\right]^{1/2}.
\ee
The propagation time to the radial boundary $R_{\rm core}$ of a CO (or He) core is
\be\label{eq:thead}
t_{\rm head} = \int^{R_{\rm core}} {dr\over v_{\rm head}(r)}.
\ee

This can be compared with the time for a spherical subset of the core, of mass $M_{\rm col}$,
to collapse and feed the engine, which we estimate to be about twice the free-fall time:
\be\label{eq:tcol}
t_{\rm col}(M_{\rm col}) \sim \pi\left[{r^3(M_{\rm col})\over 2GM_{\rm col}}\right]^{1/2}.
\ee
The formation of a black hole depends on the collapse of at least $\sim 2.5-3\,M_\odot$ of material,
and the corresponding collapse time is $\sim 0.5-3$ s (Figure \ref{fig:tcol}).  
Propagation of the jet to the stellar surface is progressively more difficult as the size of the
progenitor increases \citep{matzner03}.

\subsubsection{Can Accretion be Significantly Lengthened \\ by Torus Spreading?}

The core collapse time is some 2-3 orders of magnitude longer than the spreading time of a rotationally supported
torus near the innermost stable circular orbit of the black hole.  This has long been inferred to suggest that accretion persists only as long as mass is 
supplied by the collapsing core \citep{paczynski98}.

A large reservoir of collapsed
material with spreading time comparable to $t_{\rm col}$ could be supplied if rotation
contributed significantly to its support {\it before} collapse.  Consider a thick, collapsed torus of radius
$R_t$, angular frequency $\Omega_t$ and spreading time $t_{\rm acc} \sim \alpha^{-1}\Omega_t^{-1}$. 
Here $\alpha$ is the usual viscosity parameter.  The torus size relative to the pre-collapse radius $R_0$
can be related to the specific angular momentum $J_0$, expressed in terms of the Keplerian angular
momentum $J_K(R_0)$,
\be
{R_t\over R_0} \sim \left[{J_0\over J_K(R_0)}\right]^2.
\ee
Setting $t_{\rm acc} \gtrsim t_{\rm col}(R_0)$ implies that
\be
{J_0\over J_K(R_0)} \gtrsim \alpha^{1/3}.
\ee
This constraint is stringent enough to call into question accretion as the source of the late
bursts of energy detected from GRBs in the X-ray band.   Section \ref{s:afterglow} presents an alternative explanation.

Our discussion here focuses on rotating massive stellar cores.  These conclusions could be signficantly
modified if a significant fraction of long GRBs were triggered by the merger of a compact star with a second star (e.g. \citealt{broderick05}).

\subsubsection{Critical Jet  Luminosity}


Given that the jet must propagate through the CO material to produce a GRB, one obtains a lower bound
on $L_{\rm j,iso}$ by setting 
\be\label{eq:lmin}
t_{\rm col}(M_{\rm col}) = t_{\rm head}.
\ee
The result is shown in Figure
\ref{fig:ljet1} versus the collapsed mass $M_{\rm col}$ for a range of progenitor (CO core) masses.  The calculation is repeated for propagation of a
jet through a He core.

One observes in Figure \ref{fig:ljet2} that short escaping pulses (duration $\lesssim 3-5$ s) imply a very high
isotropic-equivalent jet luminosity, $L_{\rm j,iso} \gtrsim 10^{53}$ erg s$^{-1}$.  Such an outflow
is consistent with the nearby, hyperluminous GRB 130427A \citep{ackermann14,maselli14}.  However, it
is not consistent with most GRBs, including those with well-defined null intervals.  


What then is the origin of lower-luminosity GRBs, which make up the majority of the observed population?  
Relativistic jet fluid that reaches the jet head and shocks will flow to the side and back from the
jet head, forming a relativistic cocoon \citep{bc89}.  That is, the cocoon surrounding the jet contains
an inner magnetized component in addition to an outer component derived from the confining medium
\citep{levinson13,bromberg14}.  

Investigations of the prompt GRB emission have typically focussed
on the jet itself:  for example, \cite{levinson13} emphasize the role of a kink instability within the jet
in thermalizing a magnetic field.   Nonetheless, if the the engine lifetime is typically shorter than $T_{90}$
in long GRBs, then a significant fraction of the jet material will reach the boundary of the star before the 
head escapes, and the magnetized cocoon will contain a significant fraction of the transmitted jet energy.
At the time of jet breakout from the star (equation (\ref{eq:thead})),
\be\label{eq:ecocoon}
{E_{\rm cocoon} \over E_{\rm jet}} \sim {t_{\rm head}\over t_{\rm col} - t_{\rm head}},
\ee
where $E_{\rm jet}$ is the energy remaining in the jet column.  


\subsection{How are Ambient Baryons Removed from the Jet?}

The Lorentz factor that a magnetized jet attains following breakout also is limited
by the entrainment of baryons from the confining medium.  The relativistic jet
is separated from the shocked, confining medium by a contact discontinuity at the
jet head (e.g. \citealt{macfadyen99,matzner03,lazzati09,bromberg12}, and references therein).  

As long as the Lorentz factor of the contact is $\Gamma_c < 1/\theta$, then the shocked baryons can flow to 
the side of the jet.  But once $\Gamma_c$ exceeds $\sim 1/\theta$, the shocked baryon 
shell is effectively stuck at the jet head \citep{waxman03}.  In the absence of a 
further instability, its Lorentz factor grows as $\Gamma_c \propto r^{1/3}$ in 
response to the flow of relativistic energy from behind.  The shell becomes
geometrically thin with respect to the causal distance $\sim r/2\Gamma^2$,
especially after the onset of radiative cooling.  It is then subject 
to a corrugation instability \citep{thompson06}.

We re-examine the fate of this corrugating baryon shell in this paper.  We argue
that the column of baryons entrained by a magnetized jet is bounded from below
by the outward radiation pressure force once most of the baryons have
drained from the jet head.  Their presence in the jet column implies a strong
interaction with the magnetized jet, and presumably strong heating of the 
magnetofluid, until the flow has expanded to a distance $\sim 2\Gamma^2 ct_{\rm eng}$.
The forward baryon shell need not remain segregated from the bulk of the
relativistic outflow, as argued recently by \cite{eichler14}.

The effective breakout of the jet is displaced outward from the boundary of
the Wolf-Rayet envelope (or neutron-rich debris cloud), and the Lorentz factor
at breakout is limited to $\Gamma_{\rm br} \sim 1/\theta$.

\subsection{GRB pulse characteristics} The spectral peak energy shows rapid hard-to-soft evolution
near the beginning of a GRB pulse, and pulses are typically narrower at higher energies
\citep{fenimore95}.  Broadening of an intrinsically narrow pulse at energies {\it below} the peak is
a consequence of emission off the axis between the observer and the engine \citep{qin05},
but pulses are also observed to narrow {\it above} the peak, at least in one BATSE channel.
Finally the energy of the spectral peak, and the energy flux at the peak, 
both decline as off-axis emission begins to dominate, with a relative scaling 
$(\omega F_\omega)_{\omega_{\rm pk}} \propto t_{\rm obs}^{-2} \propto \omega_{\rm pk}^2 - \omega_{\rm pk}^{2.5}$
\citep{borgonovo01,ghirlanda10}.  Here $t_{\rm obs}$ is the observer's time and $\hbar\omega_{\rm pk}$
the peak energy.  This contrasts with the {\it asymptotic} scalings $\sim t_{\rm obs}^{-3}$
and $\omega_{\rm pk}^3$ for optically thin emission from a spherical, relativistic shock \citep{kumar00}. 

Some attention has been given to explaining this apparent discrepency by fundamentally changing the process of pulse
formation.  The decay in the spectral peak could be due to synchrotron cooling of relativistic particles \citep{preece14},
but the passive cooling rate is much too high unless the particles are embedded in a dynamically weak magnetic field.  
One could also consider continuing emission from the shell (e.g. \citealt{asano11}), but
this emission would need to continue over something like a decade in radius, given the range covered by the observed scalings.
It would give much broader pulses below the spectral peak only in the slow-cooling regime.

However, it is first important to check the scalings using a more complete Monte Carlo evaluation, which
also allows us to incorporate the effect of photospheric scattering.  We show that the asymptotic scalings
do not apply during the first part of pulse decay, due to the finite width of the emitting shell.  The
detailed result ends up being closer to the data.  Residual scattering by cold frozen pairs has the effect 
of slowing down the decay of $\omega_{\rm pk}$ and so steepening the relation between peak flux and peak frequency.  
We also consider the effects of non-spherical shell curvature and/or latitudinal gradients in $\Gamma$.  

Much broader pulses at higher energies are shown to result if the high-energy spectral tail forms by
multiple scattering (the GRB analog of an accretion disk corona:  e.g. \citealt{giannios06b,lazzati10}),
because the hard photons are created last.  By contrast,
the radiation mechanism considered by Gill \& Thompson (2014) generates the hardest photons first, and is
broadly consistent with the data.  (One must distinguish here between hard photons generated internally
to the outflow, and those inverse-Compton scattered at the forward shock.  In the second case,
there is a delay in the onset of the hard photons due to the end of pre-acceleration of
the external medium by the radiation force acting on pairs:  \citealt{thompson06,beloborodov13b}.)

\section{Interaction with Baryonic Material:  Embedding and Decoupling}\label{s:baryon}



We argued in Sections \ref{s:null}-\ref{s:magcocoon} that many long GRBs, especially those containing multiple pulses separated
by periods of low flux, are powered not directly by the jet but instead by the magnetized cocoon that forms around it.  
The isotropic luminosity that is required to punch through a CO core in the collapse time exceeds that observed in many long GRBs.

In this case, the effective radius at which the magnetized material can begin free expansion
is limited by the corrugation of a forward baryon shell, and by the subsequent
drift of baryonic fragments backward through the flow.  This corresponds to
a relatively large mass of baryons, enough to strongly perturb and heat the magnetic field.  Within an outflow of duration $\Delta t$,  
this drift is completed at a radius
\be\label{eq:rbreak}
R_{\rm br} = {\cal R}_{\rm br} \cdot 2\Gamma_{\rm br}^2 c\Delta t,
\ee
where ${\cal R}_{\rm br} \gtrsim 1$.  Here the Lorentz factor of the magnetized material at breakout is limited
to $\Gamma_{\rm br} \sim 1/\theta_{\rm j}$, and we expect that $\Delta t \sim t_{\rm eng}$, the engine active period.

In this section we first consider the geometry of the outflow and
the growth of the corrugation mode.   We then estimate
the mass of baryons that is entrained in the magnetic field by radiation 
pressure, and finally the delayed interaction between baryons and
magnetic field once the radiation compactness drops below a critical level.  
We examine the implications for burst variability in Section \ref{s:vary}.

\begin{figure}
\includegraphics[width=0.45\textwidth]{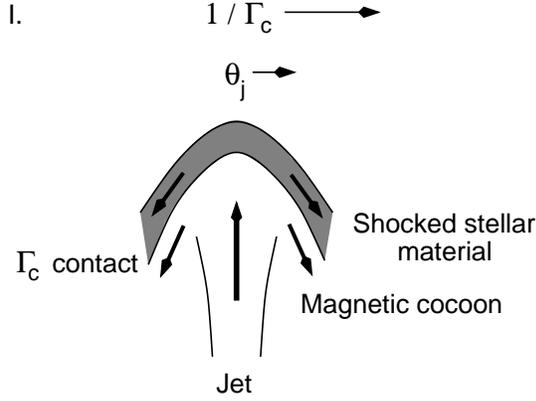}
\caption{I. Early stages of jet breakout from a confining medium.  The jet head moves slowly enough 
($\Gamma_c < 1/\theta_{\rm j}$, where $\theta_{\rm j}$ is the opening angle) that shocked stellar material can flow to the side.}
\label{fig:slip}
\end{figure}
\begin{figure}
\includegraphics[width=0.45\textwidth]{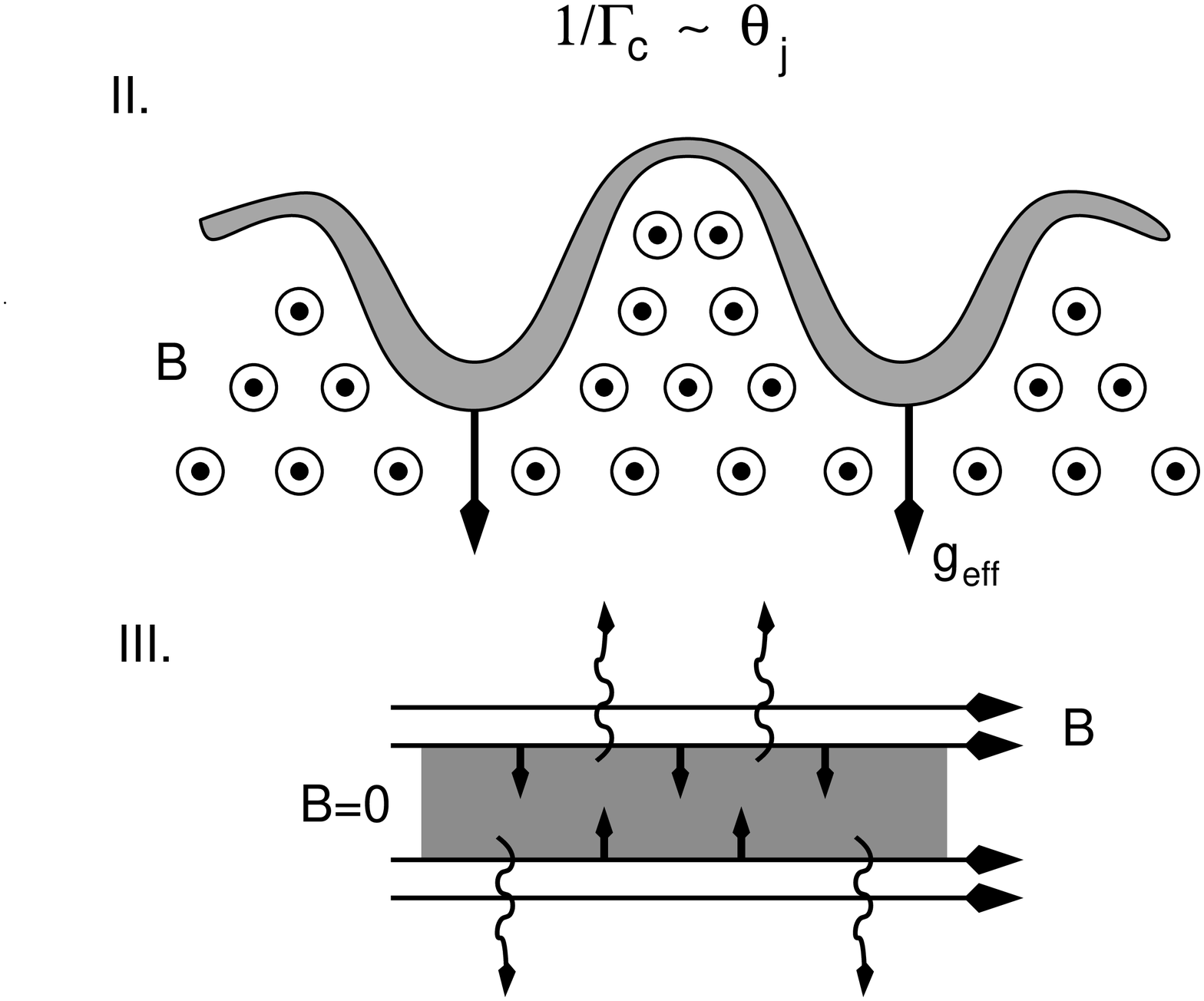}
\caption{II. Later stage of breakout:  the contact has accelerated to $\Gamma_c \sim 1/\theta_{\rm j} \gtrsim 3$ and expanded so that the
trapped baryon shell is geometrically thin.  A growing corrugation mode (${\bf k} \perp {\bf B}$) is driven by the effective
gravity (\ref{eq:geff}).  The shell is radiation-pressure dominated, and when formed is thick enough to trap the radiation.
III.~Thinning out of the corrugated shell.  The hydromagnetic instability reduces the baryon column at the jet head.  The
trapped radiation is squeezed out by the external magnetic pressure when the shell thickness drops below the critical value (\ref{eq:taumin}).}
\vskip .2in
\label{fig:corrugation}
\end{figure}
\begin{figure}
\includegraphics[width=0.45\textwidth]{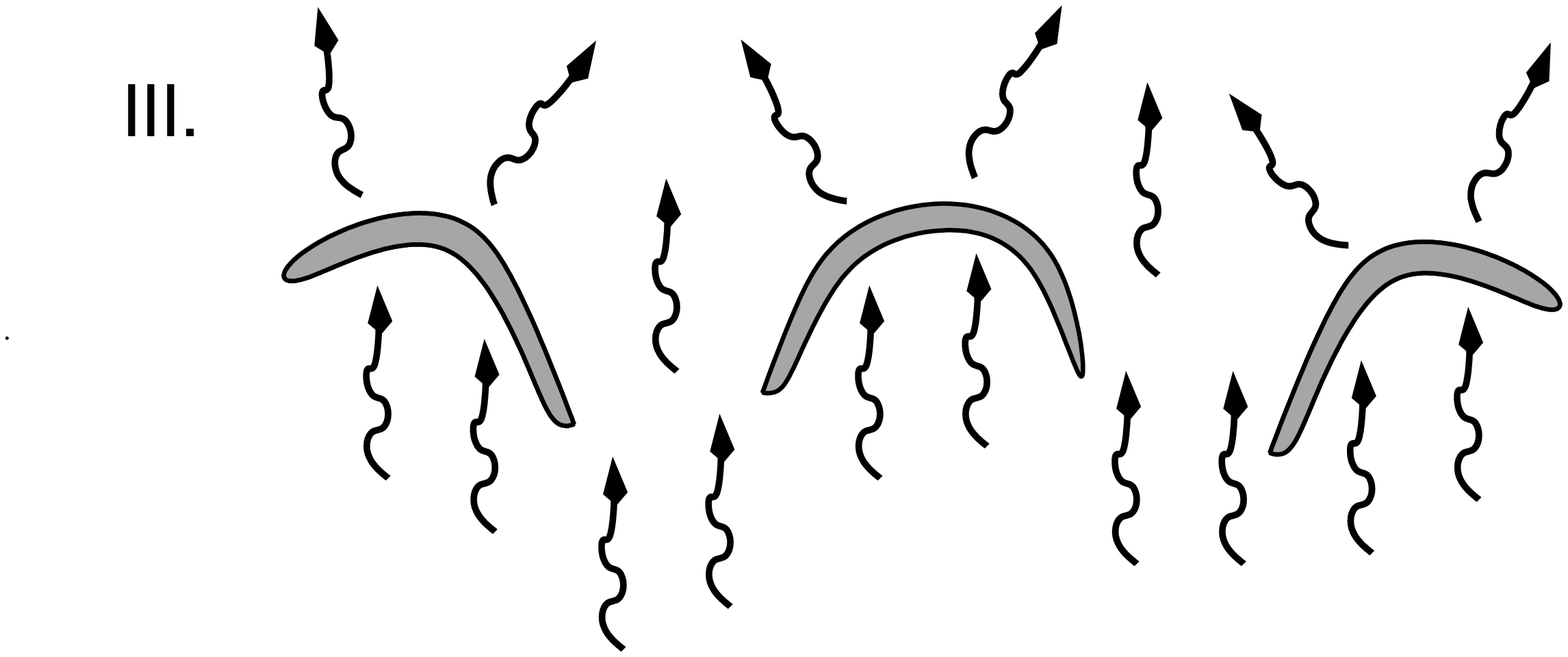}
\caption{IV. Compression by the magnetic field greatly shortens the radiative diffusion time across a
baryon shell.  Shells thin enough to cool in this way can couple effectively to the hyper-Eddington radiation flow, and 
are carried outward with the magnetic field.  The corresponding drop in baryon loading allows the magnetized material to reach
$\Gamma > 1/\theta_{\rm j}$.   Most of the baryons that experience the corrugation instability are left behind in the jet column,
and their decoupling from the magnetic field may be delayed to a radius as large as (\ref{eq:rbreak}).} 
\label{fig:entrainment}
\vskip .2in
\end{figure}

\subsection{Outflow Geometry}

We consider two simple geometrical models of a forward baryon shell.
In the `jet' geometry, the shell has Lorentz factor $\sim 1/\theta_{\rm j}$ at initial breakout from the confining medium
(radius $R_{\rm env}$).   Here the duration $t_{\rm eng}$ of the engine is longer than the breakout time $t_{\rm head}$, so the
shell continues to receive momentum from behind.  Its Lorentz factor grows
as $\Gamma_{\rm sh}(r) \sim \theta_{\rm j}^{-1} (r/R_{\rm env})^{1/3}$ until either a corrugation instability is triggered, or
most of the relativistic material has been swept up by the shell.  The second condition corresponds to 
\be\label{eq:rcause}
R_{\rm sweep} \sim R_{\rm env}\left[{2c(t_{\rm eng}-t_{\rm head})\over 3\theta_{\rm j}^2R_{\rm env}}\right]^3.
\ee
However a corrugation instability is typically triggered before this point, because the shell
cools radiatively and becomes geometrically thin \citep{thompson06}.  
More general acceleration profiles of a foward baryon shell, neglecting the effect of fragmentation, 
have been considered by \cite{eichler14}.

In the second `causal slab' geometry, energy is deposited within a relativistic (magnetized) cocoon inside $R_{\rm env}$ over a lateral size
$\theta_c R_{\rm env}$.  This relativistic material drives a baryon shell outward, which starts with $\Gamma < \theta_c^{-1}$
and initially is able to flow sideways.  The fraction of the relativistic material that remains in causal contact with the baryon shell
during breakout depends in a non-linear way on the remaining surface density $\Sigma_{\rm sh}$.   Here we simply note that
free expansion of the relativistic material allows its Lorentz factor to grow to $\sim \theta_c^{-1}$ a radius $r-R_{\rm env} 
\sim R_{\rm env}$.  Thereafter most of the relativistic material quickly catches up with the baryon shell.  We therefore
estimate a breakout Lorentz factor $\Gamma_{\rm br} \sim 1/\theta_c$.  

It is worth emphasizing that both of these geometries can be realized in
a single GRB event.  Observation of the jet emission from a source at
a fixed distance is rarer by a geometrical factor $(\theta_j/\theta_c)^2$.
On the other hand, the jet component is detectable to a greater luminosity 
distance $d_{\rm L, jet} \sim (E_{\rm jet}/E_{\rm cocoon})^{1/2}
(\theta_j/\theta_c)^{-1}\,d_{\rm L, cocoon}$.  
Therefore if the net energies carried by the two components are comparable,
$E_{\rm jet} \sim E_{\rm cocoon}$, then the jet should be dominant in very bright
GRBs above a fixed flux threshold (high enough that all the sources are
in the local Euclidean volume at cosmic redshift $z \lesssim 1$).  
Whereas the strongly-magnetized cocoon should begin to dominate in 
a complete sample of GRBs observed mostly at $z > 1$.

\subsection{Onset of Corrugation Instability}\label{s:corrug}

Outward acceleration of the magnetized jet material below the contact introduces an effective
gravity in the frame of the contact,
\be\label{eq:geff}
g_{\rm eff} = {dv_r\over dt}\biggr|_{\rm comoving} = -c^2{d\Gamma\over dr}.
\ee
When the material above the contact is much thinner than $\sim r/\Gamma$, it becomes subject to
a corrugation instability, in direct analogy with cooling gas shells behind non-relativistic
shocks (\citealt{vishniac83}, see also \citealt{duffell14}).  A horizontal magnetic field tends to suppress the growth of corrugation 
modes with wavevector ${\bf k} \parallel {\bf B}$, but not with ${\bf k} \perp {\bf B}$.  

The pressure of the forward baryon shell is provided mainly by X-ray blackbody photons, which are initially trapped in it.  
Under adiabatic expansion in response to the pressure exerted by the magnetized component, 
${B'}^2/8\pi \sim L_{\rm iso}/4\pi(\Gamma_{\rm sh} r)^2c$.  The comoving shell thickness scales as
\be\label{eq:rshcomp}
\Delta_{\rm sh}' \sim r^{-2} (r^2\Gamma_{\rm sh}^2)^{3/4}  \propto 
\left\{
\begin{array}{ll}
{\rm const} & ({\rm jet}),\\
r^{-1/2} & ({\rm causal~slab}).
\end{array}
\right.
\ee

The shell becomes corrugation unstable when it becomes geometrically thin.  Consider non-radial
corrugations of the entire shell with wavenumber $1/r \ll k \ll 1/\Delta_{\rm sh}' \ll 1$.  Considering
${\bf k} \perp {\bf B}$, so that the effect of magnetic tension can be neglected, the comoving growth 
rate of the corrugation instability is \citep{vishniac83}
\be
\gamma_{\rm corr}' \sim \left({{B'}^2\over 8\pi \Sigma_{\rm sh}} k\right)^{1/2}.
\ee
The growth time of a mode with $k\Delta_{\rm sh}' \sim 1$, normalized to the flow time $\sim r/\Gamma_{\rm sh} c$, is
\be\label{eq:corrug2}
\left({\gamma_{\rm corr}' r\over \Gamma_{\rm sh} c}\right)^{-1} \propto 
{\Gamma_{\rm sh}^2 {\Delta_{\rm sh}'}^{1/2}\over r} \propto
\left\{
\begin{array}{ll}
r^{-1/3} & ({\rm jet}),\\
r^{-5/4} & ({\rm causal~slab}).
\end{array}
\right.
\ee
One sees that the corrugation instability develops only slowly during the 
jet phase, at least before the shell is able to cool.  The development
is much faster after the outflow has made a transition from the jet 
to the causal slab geometry.

\subsection{Embedding of Baryons from the \\ Shocked Confining Medium}

The radiation flux near the breakout of a GRB jet remains orders of magnitude above the Eddington flux.
By the same token, the baryons that collect at the jet head are very optically thick and feel only
a surface radiation force.  This means that the radiation field initially does not suppress a corrugation
instability (e.g. \citealt{jiang13}).  

The non-linear effect of the corrugation instability described in Section \ref{s:corrug} 
is to thin out the baryonic shell, and displace most of it backward in the jet (Figure \ref{fig:corrugation}).  
Eventually the baryon column is sufficiently reduced that radiation can diffuse through a baryon shell 
in less than the radial flow time,
\be
t_{\rm diff} \sim \kappa_{\rm es} \Sigma_{\rm sh} {\Delta_{\rm sh}'\over c} < {r\over\Gamma_{\rm sh} c}.
\ee
At this point, radiation incident from below can flow through the shell, and the remaining baryons are
trapped in the flow.  

In the causal slab geometry, radiative cooling provides a feedback mechanism 
that regulates the fraction of the jet energy flux that is carried by 
baryons.
Dissipation between the baryons and magnetic field is then delayed to larger
radius, after the radiation force drops below a critical value.

Cooling plays a somewhat different role in the jet geometry: it provides
an effective trigger for the corrugation instability, and regulates the 
{\it compactness} at which relative motion of baryons and magnetic field
develops.  

We consider each of these cases in turn.

\subsubsection{Baryon Mass Flux in the Causal Slab Geometry}

Here we evaluate the remnant mass of trapped baryons at the cooling transition, corresponding
to the breakout radius $R_{\rm br}$ as given by equation (\ref{eq:rbreak}).  
The shocked shell and the warm magnetofluid are in near pressure balance,
\be\label{eq:rhosh}
{\Gamma_{\rm br}\over 3}\rho_{\rm ion}'c^2 \sim {{B'}^2\over 8\pi}
\ee
in the frame of the contact.  Normalizing the thickness $\Delta_{\rm sh}'$ of the shell
to the causal distance, one finds a large scattering depth across it:
\ba
\tau_{\rm T} &\sim& {3\over \Gamma_{\rm br}} {m_eY_e\over m_p}\ell_{\rm P,br} 
            \left({\Delta_{\rm sh}'\over R_{\rm br}/\Gamma_{\rm br}}\right)\nn
       &=& 3\times 10^5\,Y_{e\,0.5}\left({\Gamma_{\rm br}\over 3}\right)^{-1}
                \left({\ell_{\rm P,br}\over 10^9}\right)
                \left({\Delta_{\rm sh}'\over R_{\rm br}/\Gamma_{\rm br}}\right).\nn  
\ea
Here $\ell_{\rm P,br} = \sigma_T (r/\Gamma_{\rm br}) {B'}^2/8\pi$ is the
compactness in the comoving frame.
The large magnitude of
$\tau_{\rm T}$ means that the energy of the shocked baryons is effectively thermalized 
and converted to blackbody radiation.  

The shell can support itself against the external magnetic pressure
as long as the radiation remains trapped in it.  But the radiation diffuses out once the
shell thickness drops below
\be
{\Delta_{\rm sh}'\over R_{\rm br}/\Gamma_{\rm br}} \sim \tau_{\rm T}^{-1} \sim \left({\Gamma_{\rm br}\over 3\ell_{\rm P,br}}
{m_p\over m_e Y_e}\right)^{1/2}.
\ee
The corresponding scattering depth across the shell is
\be\label{eq:taumin}
\tau_{\rm T,cool} \sim 
5\times 10^2\,\left({\ell_{\rm P,br}\over 10^9}\right)^{1/2}
{Y_{e\,0.5}^{1/2}\over (\Gamma_{\rm br}/3)^{1/2}}.
\ee

Given this characteristic column density, the cumulative energy that is carried by the entrained baryons 
depends on the net solid angle that they cover.  This could, in principle, exceed the
solid angle of the jet.  We parameterize it by a covering factor $f_{\rm cover}$, which is normalized
so that the average rest-mass luminosity of the entrained baryons is
\be
{\langle dL_{\rm rest}/d\Omega \rangle\over dL_P/d\Omega}\biggr|_{\rm br} = 
             {f_{\rm cover}\cdot R_{\rm br}^2 \Sigma_b c^2 \over \Delta t\, dL_P/d\Omega}
\ee
at breakout.  This works out to
\be\label{eq:fbar}
{\langle dL_{\rm rest}/d\Omega \rangle\over dL_P/d\Omega}\biggr|_{\rm br} = 
            {3f_{\rm cover}{\cal R}_{\rm br}\over 2\Gamma_{\rm br}^2\tau_{\rm T,cool}}.
\ee
The last factor in this expression is the causal distance at breakout, compared
with the total radial width of the magnetized jet.

\subsubsection{Corrugation Instability and \\ Heating in the Jet Geometry}

Although we have shown that the threshold for the corrugation instability is approached more slowly
in a jet geometry (equation (\ref{eq:corrug2})), the extent of the mixing of baryons and magnetofluid
near the head of a collimated jet remains an open question.  Here we assume that the corrugation
mode of the adiabatically evolved forward shell develops slowly, and then consider the consequences
of a delayed cooling transition. 

The ratio of $t_{\rm diff}$ to the comoving flow time scales as
\be\label{eq:tratio}
{t_{\rm diff}\over r/\Gamma_{\rm sh} c} \propto \left({r\over R_{\rm env}}\right)^{-8/3},
\ee
given the scaling (\ref{eq:rshcomp}) for the shell thickness.  Therefore radiative cooling of the shell
generally sets in before the shell sweeps up all of the jet fluid.  When expression (\ref{eq:tratio}) 
is evaluated at the radius (\ref{eq:rcause}), one finds that it has a very strong (8th power) dependence on 
the small parameter $\theta_{\rm j}^2 R_{\rm env}/c(t_{\rm eng}-t_{\rm head})$.  

The scattering depth at breakout is determined by the balance 
\be
{L_{\rm j,iso}\over 4\pi R_{\rm env}^2} {R_{\rm env}\over 2\Gamma_{\rm br}^2 c} \sim \Gamma_{\rm br}\Sigma_{\rm sh} c^2,
\ee
where $\Gamma_{\rm br} \sim 1/\theta_{\rm j}$.  Then
\be
\tau_{\rm T} = {Y_e\sigma_T \Sigma_{\rm sh}\over m_p} = 
\left({r\over R_{\rm env}}\right)^{-2}\tau_{\rm T,br} \sim {Y_em_e\over m_p} \ell_{\rm j,br}\left({r\over R_{\rm env}}\right)^{-2}.
\ee
Outside breakout, the compactness drops to
\be
\ell_{\rm j} = \ell_{\rm j,br}\left({\Gamma\over \Gamma_{\rm br}}\right)^{-3}\left({r\over R_{\rm env}}\right)^{-1} = 
              \ell_{\rm j,br}\left({r\over R_{\rm env}}\right)^{-2}.
\ee
Given an aspect ratio $\Delta_{\rm sh}' = \varepsilon_{\rm sh} R_{\rm env}/\Gamma_{\rm br}$, 
the threshold $t_{\rm diff} < r/\Gamma_{\rm sh} c$ is reached at a radius
\be
r \sim R_{\rm env}\,\left(\varepsilon_{\rm sh}\,\tau_{\rm T,br}\right)^{3/8},
\ee
where the scattering depth through the forward shell has dropped to
\be
\tau_{\rm T} \sim \left({Y_em_e\over \varepsilon_{\rm sh}^3\,m_p}\ell_{\rm j,br}\right)^{1/4},
\ee               
and the jet compactness is
\ba
\ell_{\rm j} = \ell_{\rm j,br}^{1/4}\,\left({m_p\over \varepsilon_{\rm sh}\,m_eY_e}\right)^{3/4} = 
                2.6\times 10^4\,\varepsilon_{\rm sh}^{-3/4} \left({\ell_{\rm j,br}\over 10^7}\right)^{1/4}.\nn
\ea
            
This is not too different from the critical compactness at which the radiation force ceases to provide strong outward
pressure on the baryons.  We conclude that delayed dissipation driven by the differential flow of baryons and magnetic field
will occur in both the jet and causal slab geometries.

\section{Variability of the Gamma-ray Emission from a Displaced, Magnetized Fireball}\label{s:vary}

\begin{figure}
\includegraphics[width=0.45\textwidth]{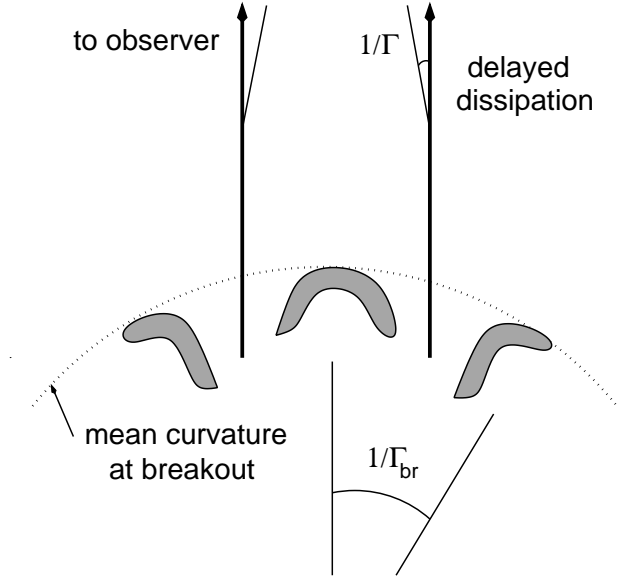}
\caption{Shocked material derived from the confining medium experiences a delayed corrugation 
instability following breakout.  Magnetized material escaping through holes in the corrugated shell
deviates from purely radial flow by an angle $\delta\theta \sim 1/\Gamma_c$.  Causal contact is then lost across an angle $\sim \theta_{\rm j}$
as the magnetized material accelerates beyond breakout.  When dissipation resumes at a larger radius $R_{\rm sat,ei}$ (equation (\ref{eq:rsat})), overlapping gamma-ray
pulses can result from causally separated events.}
\label{fig:breakout}
\vskip .2in
\end{figure}

We now consider the time dependence of the radiation from a relativistic magnetofluid
that interacts with denser baryonic material.  The starting point is the delayed breakout of
hot magnetofluid from a thin, curved shell as described in Sections \ref{s:null} and \ref{s:baryon}.
Following breakout, the embedded pairs largely annihilate and the thermal radiation field self-collimates
(Paper I).  The magnetofluid is then accelerated outward by a combination of radiation pressure and the Lorentz
force (\citealt{tchek10,russo13a,russo13b}, Paper II).  Streams formed by independent breakout events can remain
causally disconnected from each other.  The basic picture is summarized in Figure \ref{fig:breakout}.

We have suggested that the duration of the prompt gamma-ray emission, as measured by $T_{90}$,
typically exceeds the outflow duration $\Delta t$, as measured by the thickness of the magnetic shell at breakout.
The corrugation instability at the head of the shell depends on a small aspect ratio
of shell thickness to causal distance, which forces a delay in its non-linear development.  A characteristic 
pulse width is then $\Delta t$.  We have drawn a connection between the $\sim 0.5-3$ s collapse time of 
the central $2.5-3\,M_\odot$ of a Wolf-Rayet core, and the peak of the pulse width distribution in long GRBs.  

The range of pulse arrival times, corresponding to $T_{90}$, can be estimated from the geometric curvature delay,
\be
T_{90} \;\sim\; {\rm min}\left({1\over \Gamma_{\rm br}^2},\theta_{\rm j}^2\right) 
{R_{\rm br}\over 2c} \;\sim\; {\rm min}\left[1, (\theta_{\rm j}\Gamma_{\rm br})^2\right]\,{\cal R}_{\rm br} \Delta t.
\ee
Here ${\cal R}_{\rm br}$ is the normalized breakout radius (\ref{eq:rbreak}).
Null periods of GRBs correspond to angular directions in which the magnetized jet fluid is not flowing
toward the observer.

A magnetized, pair-loaded outflow has a scattering depth $\tau_{\rm T} \sim 3$ at breakout.\footnote{This is where
pair annihilation freezes out and passive expansion takes over \citep{tg13}.}  Variability is therefore concentrated on a timescale
\be\label{eq:tvarbr}
\delta t_{\rm var,br} \sim \Delta t.
\ee
Faster variations that might be generated by jet instabilities during breakout would 
be smoothed out by multiple electron scattering.  

In the following three sections, we revisit the role of radial inhomogeneities outside breakout in driving 
pulse structure at higher energies.  We also consider observational tests of our reheating mechanism, and comment 
on the role of anisotropic emission in driving variability.

\subsection{Radial Variability after Free Expansion}

The curvature delay experiences a contraction outside breakout,
but this contraction is weaker than in earlier fireball models where free expansion begins near the engine.
Therefore emission at energies above the spectral peak naturally varies on a timescale
$\delta t_{\rm var} \ll \Delta t$.  As we now show, it is still consistent with sub-pulse variability over 
a fraction $\sim (0.03-0.1)$ of a pulse width.

Given a shell of radius of curvature $R_c$ at a distance $r$ from the engine, one has (e.g. \citealt{sari97})
\be\label{eq:tdelay}
\delta t_{\rm var} \sim t_{\rm curve} \sim {r^2\theta^2\over 2R_c c} \sim {r^2\over 2\Gamma^2R_cc},
\ee
where $\theta$ is the angle between an emitting patch and the point closest to the observer.  

Considering the simplest case of free radial expansion, with Lorentz 
factor $\Gamma = \Gamma_{\rm br} (r/R_{\rm br})$, the curvature of a
reheating shell at $r \gg R_{\rm br}$ is simply $R_c = r$.
Then equation (\ref{eq:tdelay}) translates into
\be
\delta t_{\rm var} \sim {{\cal R}_{\rm br}\over \Gamma/\Gamma_{\rm br}} \Delta t.
\ee

The Lorentz factor in the dissipation zone covers the range
\be
\Gamma_{\rm sat,ei} < \Gamma < \Gamma_{\rm sat} \sim (8-10)\Gamma_{\rm sat,ei}.
\ee
Here $\Gamma_{\rm sat,ei}$ and $\Gamma_{\rm sat}$ are the Lorentz factors 
at which ions, and subsequently the magnetized pair fluid, decouple from the radiation field,  
\ba\label{eq:gamsat}
\Gamma_{\rm sat} &\sim& \Gamma_{\rm br} \left(\ell_{\gamma\,\rm br}\right)^{1/4}\nn
                &=& 300\,\left({\ell_{\rm \gamma,br}\over 10^8}\right)^{1/4}
            \left({\Gamma_{\rm br}\over 3}\right);\nn
\Gamma_{\rm sat,ei} &=& \left({Y_e m_e\over m_p}\right)^{1/4}\Gamma_{\rm sat} = 0.13\,Y_{e\,0.5}^{1/4}\Gamma_{\rm sat},
\ea
and the corresponding radii are
\ba\label{eq:rsat}
R_{\rm sat} &=& {\Gamma_{\rm sat}\over\Gamma_{\rm br}}\,R_{\rm br}\nn
            &=& 5.4\times 10^{14}\,\left({{\cal R}_{\rm br} \Delta t\over 10~{\rm s}}\right) \left({\ell_{\rm th,br}\over 10^8}\right)^{1/4}
             \left({\Gamma_{\rm br}\over 3}\right)^2\quad {\rm cm},\nn
R_{\rm sat,ei} &\sim& 0.13\,Y_{e\,0.5}^{1/4} R_{\rm sat}.
\ea
Here the breakout compactness 
\be\label{eq:lbreak}
\ell_{\gamma,{\rm br}} = {3\sigma_T E_{\rm \gamma,iso}\over 32\pi {\cal R}_{\rm br}\Gamma_{\rm br}^5 m_ec^4\Delta t^2}
\ee
is expressed in terms of the isotropic breakout (thermal) photon energy.

Consider, for example, the case where the Lorentz factor in the baryon-free parts of the outflow has grown to
$\Gamma \sim 2\Gamma_{\rm sat,ei}$ (and the photon compactness has dropped to $\ell_\gamma \sim 2^{-4}
m_p/Y_em_e \sim 230$).  Then 
\be\label{eq:tvary}
\delta t_{\rm var} \sim 0.4\,
{({\cal R}_{\rm br}/10)^{5/4}\over (E_{\rm\gamma,iso}/10^{52}~{\rm erg})^{1/4}}
\left({\Delta t\over {\rm s}}\right)^{3/2}\left({\Gamma_{\rm br}\over 3}\right)^{5/4}\;{\rm s}.
\ee
This allows high-energy photons that are generated during reheating to vary
on a somewhat shorter timescale than the spectral peak.

The estimate (\ref{eq:tvary}) depends on our estimate $R_c \sim r$ 
in the delayed dissipation zone, corresponding to a conical outflow.  
A smaller radius of curvature (and therefore larger $\delta t_{\rm var}$) 
is obtained if the outflow is more collimated.  Taking for example
$\theta \sim r^{-\zeta}$ along a streamline corresponds to 
$R_c = r(1-2\zeta)/(1-\zeta)^2$.  The correction ends up being modest
except in the case of extreme (e.g. parabolic) collimation.

\subsection{Variability due to Delayed Magnetic Reconnection?}\label{s:rec}

Variability in the outflow from a black hole engine could arise from a dynamo process operating in a surrounding torus
\citep{thompson06}.  The reversal time $t_{\rm rev}$ of the magnetic flux threading the event horizon may greatly exceed the 
orbital period $P_{\rm orb} \sim 10^{-3}$ s at the inner boundary of the torus, 
with the result that magnetic reconnection in the outflow is delayed to a considerable radius.  Fast reconnection also typically
takes place at a fraction of the Alfv\'en speed.  Taking $t_{\rm rev} \sim 10^2 P_{\rm orb} \sim 0.1$~s and a reconnection
speed $V_{\rm rec} \sim 0.1 c$, one finds that the advected magnetic energy begins to dissipate only after the outflow
has propagated for a time
\be
t_{\rm rec} \sim \Gamma^2 {c t_{\rm rev}\over V_{\rm rec}} \sim 10~\left({\Gamma\over 3}\right)^2\left({t_{\rm rev}\over 0.1~{\rm s}}\right)
\quad {\rm s}
\ee
in the inertial frame.  Note that this timescale is {\it longer} than the $2.5$-$3 M_\odot$ collapse time in Wolf-Rayet
cores (Figure \ref{fig:tcol}), as well as our posited $\Delta t \sim 1$ s outflow duration.   

For this reason, the importance of magnetic reconnection in powering dissipation in magnetized GRB outflows is quite uncertain.
It would have a greater importance if the outflow accelerated more slowly {\it and} the striping had a shorter lengthscale
\citep{spruit01,giannios06c,mckinney12}.

\subsection{Anisotropic Emission in the Comoving Frame?}

Variability on timescales shorter than (\ref{eq:tvary}) is certainly
observed in some long GRBs.  

The emission process described in \cite{thompson06} and Paper II is 
anisotropic, due to longitudinal heating of the pair gas along the magnetic 
field.  This introduces a mechanism for generating narrower
pulses at higher photon energies, which are emitted by more relativistic
$e^\pm$ with narrower Lorentz cones.  

This effect is washed out, in part, by gradients 
in the field direction across the heating zone, combined with rescattering 
by the regenerated pairs ($\tau_{\rm es} \sim 1$-4 at the end of reheating).
Nonetheless, since higher energy photons are emitted first
during pair breakdown, some imprint of anisotropic emission may survive at
higher energies, especially 
when $\tau_{\rm es}$ lies closer to unity.

Because we find lower $\tau_{\rm es}$ in bursts with softer high-energy
spectra (Paper II), there is an interesting test of anisotropic emission here:
very fast variability (compared with (\ref{eq:tvary})) should be 
found most commonly
in bursts with intermediate high-energy photon indices $\beta$.  
We expect that high-energy pulses are more strongly smeared when $\beta$ 
is close to $-2$; whereas the high-energy variability is suppressed in
cases of weak reheating (a thermal spectral cutoff).

Bulk relativistic motion that is driven by magnetic reconnection would 
have a related effect \citep{lyutikov03,narayan09,lazar09}.  Localized 
flows of this type have an angular scale smaller than $1/\bar\Gamma$, 
where $\bar\Gamma$ is the Lorentz factor of the background flow.  If
such a structure were responsible for the formation an entire pulse,
then the peak of the emission would not followed by significant off-axis 
emission.  The decay of $\omega F_\omega$ at the spectral peak would
be much sharper than is generally observed.

\section{Implications for X-ray Flares \\ in Early Afterglow}\label{s:afterglow}

Discrete X-ray flares are sometimes observed superposed on the declining non-thermal afterglow in an interval $(1-10^3)T_{90}$
following a GRB (\citealt{falcone07,swenson14,margutti11}).  The corresponding delays are $10^2-10^4$ s
for long GRBs and $\sim 1-10^2$ s for short GRBs.   They appear to be closely related to the phenomenon
that produces pulses in the prompt emission.  

Here we examine how X-ray flares can result from a range of baryon loadings 
in the magnetized ejecta.  The source material expands non-relativistically, but it differs from existing cocoon models (e.g. \citealt{rr02}) in containing
a significant buried relativistic component.

The X-ray flares show hard-to-soft evolution similar to pulses in the prompt phase, but
downscaled in peak energy \citep{chincarini10,margutti10}.  Importantly the occurance of even a bright spike does not seem
to perturb the extended, nonthermal flux curve \citep{falcone06}.  The emitting material therefore appears to be offset radially from
the forward shock, either because it has a lower Lorentz factor, or it was ejected later from the engine.  If powered
by collisions between ejecta shells, the observed narrowness of the flares supports later ejection, and implies a long extension
of the accretion phase onto the black hole (e.g. \citealt{lazzati07}).  

A delayed magnetic Rayleigh-Taylor instability can circumvent the conclusion that extended accretion is taking place.  First consider the emission
of a second shell that is nearly as relativistic as the shell that produces the initial gamma-ray burst.
The electromagnetic energy (\ref{eq:emag}) that it carries is preserved only
to the extent that the shell does not spread radially: otherwise it
scales as $E_P \propto \Delta r^{-1}$, as can be seen from the conservation
of toroidal magnetic flux.  The key point here as that a lengthening of 
the pulse duration depends on radial spreading, $\Delta t_X \sim \Delta r/c$, 
and so implies a significant conversion of magnetic energy to bulk kinetic 
energy.

In the case of the giant X-ray flare of GRB 050502B, 
which peaked $t_X \sim 800$ s after the burst trigger and with a FWHM $\Delta t_X/t_X \sim 1/4$, the pulse
width was about $\sim 10^2$ times longer than the main gamma-ray pulse \citep{falcone06}.  This degree of
broadening would reduce $E_P$ by a factor $\sim 10^{-2}$.  But the 
X-ray output of the flare following GRB 050502B was even larger than the bolometric gamma-ray output.  
A similar (or larger) aspect ratio is encountered in other X-ray flares, and so we conclude that
a fully relativistic, magnetized shell is not a promising approach to the origin of the X-ray flares.  

\cite{giannios06a} has suggested that X-ray flares arise from a delayed reconnection instability in the same magnetized shell that
produced the prompt gamma-rays.  Such a shell could indeed continue to dissipate as it decelerated passively against
the ambient medium, but only at the expense of a significant loss of energy due to radial spreading.

Now let us consider an outflow containing a combination of relativistic magnetofluid and shocked stellar material.
Raising the {\it net} proportion of baryons does not necessarily reduce the extreme magnetization of the relativistic
component.  This depends on the degree of small-scale mixing, which we have
argued (Section \ref{s:baryon}) is 
delayed well beyond the initial breakout.
Consider such a shell that is ejected with speed $V_{\rm ej} < c$, kinetic energy $E_k$, and a characteristic 
width $\sim R_{\rm env}$.  Ejection is partly facilitated by a shock that accelerates outward below the stellar surface
(e.g. \citealt{tan01}), which means that $V_{\rm ej}$ can exceed the internal Alfv\'en speed that is constructed
from the r.m.s. magnetic field and the inertia of the entrained baryons,
\be
V_{\rm ej} > V_{\rm A} = \left({E_P\over E_k}\right)^{1/2} V_{\rm ej}.
\ee

A corrugation instability of the forward
baryon shell is delayed for the same reason as in the relativistic case:   its relative thickness 
decreases with radius.  Consider an initial state for the shell where the shocked baryons dominate its volume.
Then the radiation pressure filling the baryons drops as $\sim r^{-8/3}$, and the magnetic field in the
embedded relativistic fluid as $r^{-4/3}$.  The proportion of the volume filled by the magnetic field rises
as $\sim r^{1/3}$, and the magnetic energy decreases slowly, $E_P \sim r^{-1/3}$.  Non-linear breakout of the magnetic field occurs
on the timescale $\Delta t_X \sim R_{\rm env}/V_{\rm A}$.  Taking $R_{\rm env} \sim 4\times 10^{10}$ cm, one requires 
$V_A \sim 10^{-2}c$ in the case of GRB 050502A.  We re-emphasize that this Alfv\'en speed represents a balance between magnetic
stresses and the inertia of the confining baryons; the internal Alfv\'en speed in the magnetized fluid may still approach $c$.

\section{Observational Imprint of Rescattering in a Curved Relativistic Shell}\label{s:shell}

The observed pulse evolution of GRBs \citep{fenimore95,borgonovo01,ryde02,ghirlanda10,preece14} resembles the 
radiative emission from curved, relativistic shells, but with some quantitative differences that have not been explained.  

First, the decay of the flux due to off-axis emission appears to be somewhat slower ($\sim t_{\rm obs}^{-2}$) than predicted by
the asymptotic theory of impulsive emission from a spherical shell ($\sim t_{\rm obs}^{-3}$:  \citealt{kumar00,dermer04}).
Second, one observes a somewhat softer dependence of the flux measured at the peak on the decaying peak energy
$(\omega F_\omega)_{\rm pk} \sim \omega_{\rm pk}^{2-2.5}$ instead of $\omega_{\rm pk}^3$).  

In this section we first reconsider the emission from a spherical shell using a simplified approach that can be generalized to other
geometries.  Spherical symmetry is no longer a natural starting point when, for example, the spectral peak is formed
by the breakout of magnetized fluid from a baryon shell, followed by a moderate expansion to a pair photosphere.
By departing from spherical curvature, and allowing for a decrease in Lorentz factor away from the point on the 
emitting shell that is closest to the observer, we obtain new scalings.

Then we calculate in detail pulse evolution in both spherical and jet geometries, using the Monte Carlo method described
in Appendix \ref{s:monte}.  We find that even in the simplest case of optically thin emission from a spherical shell, the initial
decay of a pulse is significantly flatter than the asymptotic theory would suggest.  Scattering is shown to have differing
effects, depending on whether the pairs are cold or hot.  The pulse broadening is only slight when the pairs are cold,
but there is a dramatic widening of pulses at higher photon energies when they are hot, in strong disagreement with the data.

\subsection{Discrete Sum over Individual Emitters}

Consider a `blob' of plasma that emits $N_\gamma$ photons of energy $\hbar\omega_{\rm pk}'$ isotropically in its rest frame.
This emission may be repeated over an extended period of time, as the blob expands away from the center of the explosion.
In this simple model, each episode of emission is associated with a particular radius $r$ and Lorentz factor $\Gamma$, corresponding
to a tranverse area $\sim (r/\Gamma)^2$ and emission time $\sim r/2\Gamma^2 c$ as seen by an observer who is positioned in 
the direction of motion of the blob.

When the blob is offset
by an angle $\theta$ from the line between the engine and the observer, the observed energy is
\be
\hbar\omega_{\rm pk}(\theta) = {\cal D}(\Gamma,\theta)\hbar\omega_{\rm pk}' \simeq {2\Gamma \hbar\omega_{\rm pk}'\over 1+(\Gamma\theta)^2}.
\ee
We evaluate the Doppler factor ${\cal D} = [\Gamma (1-\beta\cos\theta)]^{-1}$ at small $\theta$ and large $\Gamma$.
A comoving emissivity $j'_{\omega_{\rm pk}'}$ corresponds to $N_\gamma \sim (4\pi/\hbar c) j'_{\omega_{\rm pk}'} (r/\Gamma)^4$.

A telescope of area $A$ at a distance $D$ subtends a solid angle $\Delta\Omega_{\rm obs} \sim A/D^2$, which is $\Delta\Omega'_{\rm obs} = 
{\cal D}^2\,\Delta\Omega_{\rm obs}$ in the frame of the blob.  The number of detected photons is, therefore,
\be
N_{\rm \gamma\,obs}(\theta) = {\Delta\Omega'_{\rm obs}\over 4\pi} N_\gamma \sim {N_\gamma\over [1+(\Gamma\theta)^2]^2} {\Gamma^2 A\over \pi D^2}
\ee
over a time 
\be
t_{\rm obs}(\theta) \sim {r\over 2\Gamma^2 c}[1+(\Gamma\theta)^2].
\ee
Then the flux is
\be\label{eq:fblob}
F_\omega(\theta) \sim {\hbar N_{\rm \gamma,obs}(\theta)\over A\,t_{\rm obs}(\theta)} = {\hbar \over 4\pi \Gamma^2}
\left({r\over D}\right)^2 {N_\gamma\over c^2 t_{\rm obs}^3(\theta)}\quad (\theta\Gamma > 1).
\ee

The simplest case is of optically thin emission that is concentrated at a particular radius $R_{\rm em}$.
Expression pre-supposes the presence of an emitting blob at angle $\theta$.  A smooth light curve extending
over a range of observer time depends on the presence of multiple such structures placed at a range of angles. 
If the ejecta have spherical symmetry, then 
\be
F_\omega^{\rm sphere}(\theta) \sim \pi (\Gamma\theta)^2 F_\omega(\theta) = {\hbar R_{\rm em}\over 2 D^2}
{N_\gamma\over ct_{\rm obs}^2(\theta)} \quad (\theta\Gamma > 1).
\ee
This time-scaling, corresponding to $\sim t_{\rm obs}^{-3}$
in $(\omega F_\omega)_{\rm pk}$, was derived by \cite{kumar00} and 
\cite{dermer04} from a direct integration of the radiative transfer equation.

\subsection{Non-Spherical Shell}\label{s:pulsescale}

The preceding review of optically thin, off-axis emission from a spherical shell is now easily generalized to a more general
geometry and Lorentz factor profile.  We assume that the shell is rotationally symmetric about the axis connecting the
observer to the closest point on the shell.\footnote{Following the discussion 
in Section \ref{s:null} and \ref{s:baryon}, this may represent a local minimum in the observer-shell separation.}
The position $z$ of the shell, measured along this axis, and the Lorentz factor
follow the scalings
\ba
z(R_\perp,r) &\;=\;& z_0(r)\left({R_\perp\over r}\right)^{\alpha};   \nn
\Gamma(R_\perp) &\;\propto\;& \left({R_\perp\over r}\right)^{-\delta} \quad\quad [\theta > 1/\Gamma(0)].
\ea
The thickness of the shell vanishes, which is consistent with (e.g.) fast
synchrotron cooling at a forward shock, but not with volumetric emission
during pair breakdown.  Here $R_\perp$ is the transverse coordinate.  

If the `explosion' is offset from the engine then, even for a spherical shell
($\alpha = 2$), the angle $\psi = dz/dR_\perp = \alpha z/R_\perp$ between the normal to the shell and the direction of the observer will not 
equal the angular displacement $\theta = R_\perp/r$ as viewed from the engine.  In general the shell velocity does not follow the 
shell normal, but to simplify the discussion here we assume such a proportionality.  Then the Doppler factor is
\be
{\cal D} \simeq {2\over \psi^2\Gamma} \propto R_\perp^{-(2\alpha-\delta-2)}
\quad (\psi\Gamma > 1).
\ee

The time delay is $t_{\rm obs} = z/c \propto R_\perp^\alpha$, and the downshifted peak energy
\be
\hbar\omega_{\rm pk}(t_{\rm obs}) = {\cal D}\,\hbar\omega_{\rm pk} \propto t_{\rm obs}^{-(2\alpha-\delta-2)/\alpha}.
\ee
Here we have taken the observed peak energy along a ray $\psi = 0$ to be proportional to $\Gamma$. 
This is appropriate if the spectral peak is set by thermalization at a characteristic comoving temperature, as
when photon creation is buffered by pair annihilation \citep{tg13}.

The peak energy flux scales as
\ba
\left(\omega F_{\omega}\right)_{\rm pk} &\propto& {{\cal D}^3 \omega_{\rm pk}'[\theta\Gamma(0)]^2 \over t_{\rm obs}}
     \propto {R_\perp^2\over z (dz/dR_\perp)^6 \Gamma^3} \nn
    &\propto & t_{\rm obs}^{-(7\alpha-3\delta-8)/\alpha} \propto \omega_{\rm pk}^{(7\alpha-3\delta-8)/(2\alpha-\delta-2)}.\nn
\ea
In the case of a spherically curved shell, one finds
\be
(\omega F_\omega)_{\rm pk} \propto t_{\rm obs}^{-(6-3\delta)/2} \propto \omega_{\rm pk}^3\quad\quad ({\rm sphere}).
\ee
This follows the optically-thin frequency dependence for any any Lorentz factor profile, but the time scaling can be softened.
 
One may consider other shell geometries to obtain the observed frequency index $2-2.5$ \citep{borgonovo01,ghirlanda10}.  For example, 
when $\alpha = 3/2$ and $\Gamma$ is uniform across the shell, one finds $(\omega F_\omega)_{\rm pk} \propto t_{\rm obs}^{-5/3} \propto \omega_{\rm pk}^{5/2}$.
The scalings flatten to $(\omega F_\omega)_{\rm pk} \propto t_{\rm obs}^{-2/3} \propto \omega_{\rm pk}^2$
when $\delta$ increases to $1/2$.  Indeed the time-scalings of flux and frequency become uncomfortably flat in the presence
of a latitudinal Lorentz factor gradient.  However, a photosphere has a strong steepening effect on the scalings, as we show 
in Section \ref{s:scatter}, which means that a latitudinal gradient may still be implicated in the case of finite scattering depth.

Finally, when $\Gamma$ varies with position across the shell, even an asymptotically conical geometry ($\alpha = 1$) allows for an increasing 
Doppler factor (and decreasing observed peak energy) with angle $\theta$:  
$\omega_{\rm pk}(t_{\rm obs}) \propto t_{\rm obs}^{2\delta}$.  But, in this case, the peak energy flux has a reasonable
scaling only if $\delta < 0$, and the scaling is sensitive to the value $\delta$.

\subsection{Effect of a Photosphere}\label{s:scatter}

The emission model developed in Papers I and II involves i) the build-up of pairs to $\tau_{\rm T} \sim 10$ during the formation of 
the spectral peak; and ii) the regeneration of pairs during the formation of the high-energy spectral tail, this time
reaching a Thomson depth $\tau_{\rm T} \sim 1$-4.  

Therefore we must consider the effect of scattering on the emission of off-axis photons.  This can be reliably accomplished with a
Monte Carlo approach during a late stage of the flow, when the pairs are passively diluted by the expansion, with modest annhilation.  

A key point is that, because the photosphere is spread out significantly in radius at $\theta > 1/\Gamma$,
the decaying tail of a GRB pulse gives a direct probe of the dynamics of the emitting material.  Variations
in $\Gamma$ can occur both radially and in angle,
due to the differential flow of magnetized material with respect to slower clumps of baryons.  

The imprint of angular variations in $\Gamma$ requires a simulation involving at least axial symmetry.
We adopt a simple parameterization of the flow profile,
\be\label{eq:gamprof}
\Gamma(r,\theta) = {\Gamma_0\over \left[1+(\Gamma_0\theta)^2\right]^{\delta/2}}.
\ee
We work in the small-angle approximation.  

Non-spherical shell curvature is neglected when considering the effects 
of scattering, so that the local flow is always in the radial direction.
We do additionally use the Monte Carlo approach to calculate 
optically thin emission from non-spherical shells of a finite thickness.

Angular variations in $\Gamma$ can lead to strong variations in scattering depth, due to the $(1-\beta) 
\sim 1/2\Gamma^2$ dependence of the scattering rate.  If the slower motion  were due to a higher baryon
loading, then photons leaving the jet core could be trapped and released at a much larger radius.  
Angular variations in variable baryon loading are expected (e.g. \citealt{rr02}), but there are
no detailed predictions of their magnitude.

Here we make use of the self-regulation of the scattering depth during pair breakdown to factor out 
the effect of angular variations in $\Gamma$ on the scattering rate.  Consider a radial flow that
is sheared in the $\theta$ direction.  A photon moving at angle $\psi$ with respect to the local
flow direction sees a scattering depth which can be written as
\be
{d\tau_{\rm T}\over dr} = (1+\Gamma^2\psi^2) {\sigma_T\,n_e(r,\theta)\over 2\Gamma^2(r,\theta)}.
\ee
Hence we take
\be
{n_e(r,\theta)\over \Gamma^2(r,\theta)} \rightarrow {n_e(R_0,0)\over \Gamma^2(r,0)}\left({r\over R_0}\right)^{-2},
\ee
where $R_0$ is a reference radius marking the peak of dissipation in 
the outflow, and $\Gamma_0$ the corresponding Lorentz factor.  The factor 
$r^{-2}$ represents outward advection of the particles after pair 
annihilation freezes out.

A photon that escapes to the observer from the reheated jet typically has experienced more than one scattering.
Its arrival time is delayed because of three geometrical effects.

Zero time is identified with a photon emitted radially from the front of the dissipating shell
at the end of dissipation (corresponding to a radius $R_0$).  The emitting shell has thickness
$\Delta R_{\rm em}$ and is labelled by a radial coordinate $0 \leq \xi_{\rm em} \leq \Delta R_{\rm em}$.  The
corresponding delay is
\be
t_{\rm shell} = {\xi_{\rm em}\over c}.
\ee
An additional geometric delay is accrued at a rate
\be
{dt_{\rm geom}\over dr} = {1\over c}\left({1\over\cos\psi} - 1\right) \simeq {\psi^2\over 2c},
\ee
due to radial drift between the photon and a radially moving reference photon.   The angle $\psi$ 
decays from the value at a previous scattering (at a radius $r_{\rm em}$) according to
\be
\psi = \psi_{\rm em} {r_{\rm em}\over r}.
\ee

The final geometric delay, incurred at the last scattering, depends on the radius of
curvature $R_c$ of the emitting shell.  In an axisymmetric jet with observer oriented
at angle $\theta_{\rm obs}$ with respect to the jet axis, this delay is defined with 
respect to the observer-engine axis.  Then a photon scattered toward the observer
at radius $r_{\rm last}$ receives the delay
\be
\Delta t_{\rm geom}^{\rm last} = \psi_{\rm last}^2 {r_{\rm last}^2\over 2 R_c c}.
\ee
The net arrival time is the sum of these three contributions, written in the order that they are
generated:  
\ba
\Delta t_{\rm obs} &=& \Delta t_{\rm shell} + \Delta t_{\rm geom} + \Delta t_{\rm geom}^{\rm last}\nn
                   &=& {\xi_{\rm em}\over c} + \Delta t_{\rm geom} +  \theta_{\rm last}^2 {r_{\rm last}^2\over 2R_cc}.
\ea

The final energy of a photon originating with $\hbar\omega$ is obtained
by a sequence of Doppler shifts,
\be
\omega  \rightarrow  \omega_{\rm em} =  {1 + (\Gamma\psi)^2 \over 1 + (\Gamma\psi_{\rm em})^2} \omega,
\ee
where $\psi,\psi_{\rm em}$ are the propagation angles of the incident and scattered photons.

\subsubsection{Effect of Residual Scattering on Output Spectrum}

\begin{figure}
\includegraphics[width=0.48\textwidth]{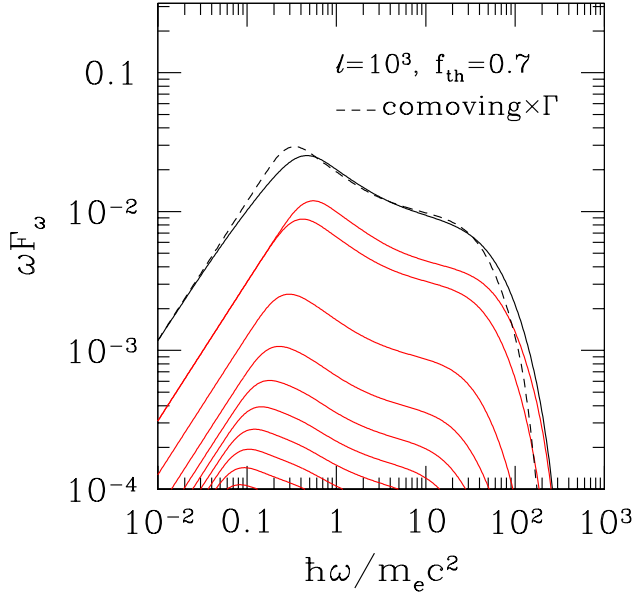}
\caption{Output spectrum (solid black curve) resulting from multiple scattering off a passively expanding
pair gas with constant Lorentz factor $\Gamma_0 = 100$ and initial scattering depth $\tau_{\rm T} = 3$, as
determined by the end of the kinetic calculation ($\ell_{\rm tot} = 10^3$ and $f_{\rm th} = \ell_{\rm th}/\ell_{\rm heat} 
= 0.7$ from Paper II).  Dashed black line:  source spectrum, boosted by a factor $\Gamma_0$.  
Red curves:  time-resolved spectrum, plotted at intervals $\Delta t = 0.5 (R_0/2\Gamma_0^2c)$.}
\label{fig:scattspec}
\vskip .2in
\end{figure}
\begin{figure}
\includegraphics[width=0.48\textwidth]{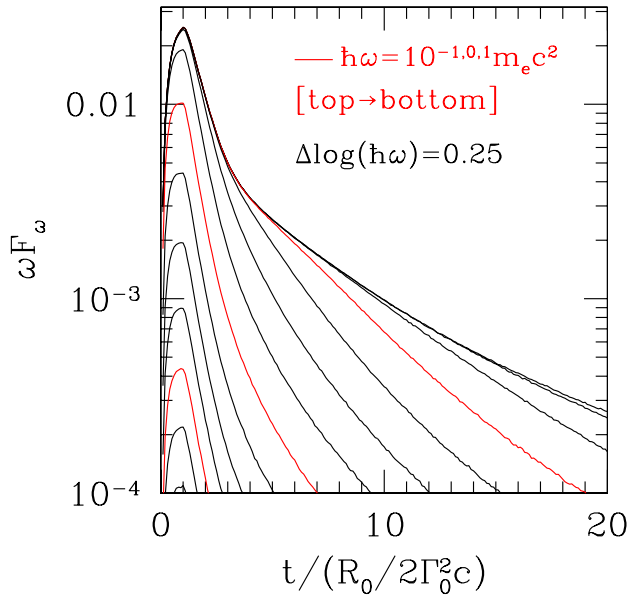}
\caption{Pulse profiles corresponding to the sequence of spectra in Figure \ref{fig:scattspec}.  Red curves
now label a subset of the snapshots, separated in time by $\Delta\log(t) = 1$.}
\label{fig:pulse}
\end{figure}
A sample photon spectrum obtained from the static kinetic calculations of Paper II is now evolved by scattering
off the frozen, expanding pair gas.  The assumption here is that dissipation is concentrated over a relatively narrow
range of radius near $R_0$.  The resultant pairs (which are cold at the end of the kinetic calculation)
are then assumed frozen into the flow.  To facilitate comparison with other calculations, we take a fixed scattering 
depth $\tau_{\rm T}(R_0) = 3$, as evaluated using the radial integral
\be\label{eq:tauT}
\tau_{\rm T}(r) = \int_r^\infty \sigma_T n_e(r) {dr\over 2\Gamma^2}.
\ee
The residual thermal Compton parameter of the pair gas, 
$y_{\rm C} \sim \tau_{\rm T} \kB T_e/m_ec^2$, here is assumed to vanish.

In Figure \ref{fig:scattspec} we compare the output spectrum, averaged over an entire pulse, with the one-box
calculation boosted by a factor $\Gamma$ in energy.  The time resolved spectrum shows
little evolution in shape, except for an overall reduction in energy due to side-ways
emission.   The corresponding pulse profiles are shown in Figure \ref{fig:pulse}.  

\subsubsection{Pulse Profiles and Spectral Cooling}\label{s:coldscatt}

Next we confront the simplest model of a dissipating shell with {\it spherical} symmetry against two
sensitive observational tests of the emission geometry in GRBs:  the dependence of pulse width
on photon energy \citep{fenimore95,qin05}; and the scaling between peak energy flux and spectral peak energy
within a burst \citep{borgonovo01,ghirlanda10}.  

\begin{figure}
\includegraphics[width=0.48\textwidth]{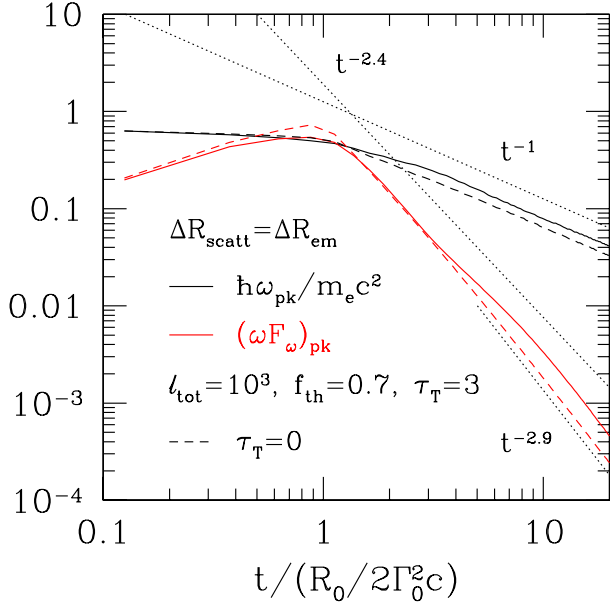}
\caption{Time variation of spectral peak energy and energy flux at the spectral peak, in a burst
based on our one-zone spectral model with $f_{\rm th} = 0.7$, compactness $\ell_{\rm tot} = 10^3$, and $\Gamma_0 = 10^2$
during the emission of the high-energy tail.  Here scattering charges are localized in the emission zone, in a shell of width
$\Delta R_{\rm scatt} = \Delta R_{\rm em} = R_0/2\Gamma_0^2c$.   Dashed lines: optically thin emission.  The decay is initially flatter than $\sim t^{-3}$ 
even in the case of optically thin emission: the asymptotic scaling only sets in about a decade below the peak.}
\label{fig:flux_peak}
\vskip .2in
\end{figure}

The decay of peak flux and peak energy due to off-axis emission is 
shown in Figure \ref{fig:flux_peak}, using the same set-up as for Figures \ref{fig:scattspec} and \ref{fig:pulse}.
A useful reference point is provided optically thin emission with the same input spectrum.
In this case, the first part of the pulse decay is significantly flatter ($\sim t_{\rm obs}^{-2.4}$) than the
asymptotic scaling (which is close to the analytic value $\sim t_{\rm obs}^{-3}$).  The addition of 
scattering has an additional flattening effect on both the late pulse tail and the decay of the peak energy. 

\begin{figure}
\includegraphics[width=0.48\textwidth]{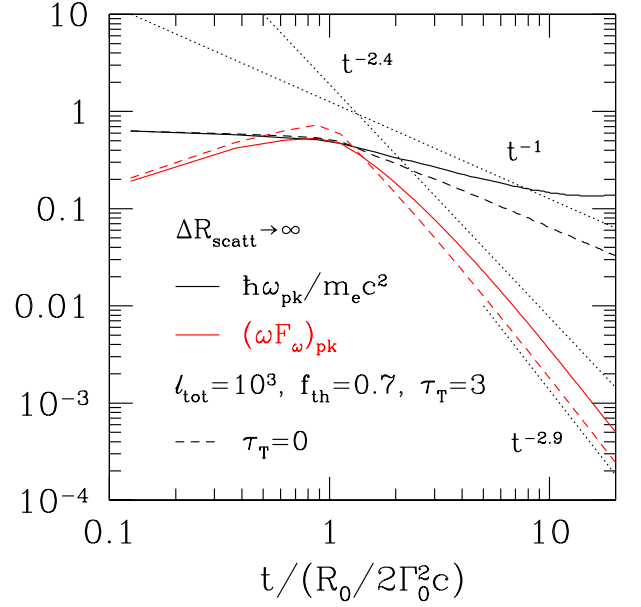}
\caption{Same as Figure \ref{fig:flux_peak}, but now the scattering particles are present far beyond the emission
shell, $\Delta R_{\rm scatt} \gg \Delta R_{\rm em} = R_0/2\Gamma_0^2 c$, as they would near the photosphere of a baryon-dominated outflow.  This
leads to a more pronounced scattering wing and a plateau in the peak energy.}
\label{fig:flux_peak2}
\vskip .2in
\end{figure}

It is also interesting to compare the effect of a cold scattering atmosphere that is localized within
the emitting shell (the choice made so far) with more extended particle flow (Figure \ref{fig:flux_peak2}).  The point here is that 
sudden emission is limited by causality to a radial shell of width $\Delta R_{\rm em} \lesssim R_0/2\Gamma_0^2$.
The scattering particles will also be localized in a shell of thickness $\Delta R_{\rm scatt} \sim \Delta R_{\rm em}$ 
if they are generated during the same re-heating episode (Paper II).
Charges generated at a smaller radius and lower $\Gamma$, or baryons frozen into the outflow, are an alternative
source of photospheric scattering.  In this second case $\Delta R_{\rm scatt} \gg \Delta R_{\rm em}$.  One observes
in Figure \ref{fig:flux_peak2} that this introduces a broad plateau in $\omega_{\rm pk}$ which generally is
inconsistent with observed GRB pulses.

\begin{figure}
\includegraphics[width=0.48\textwidth]{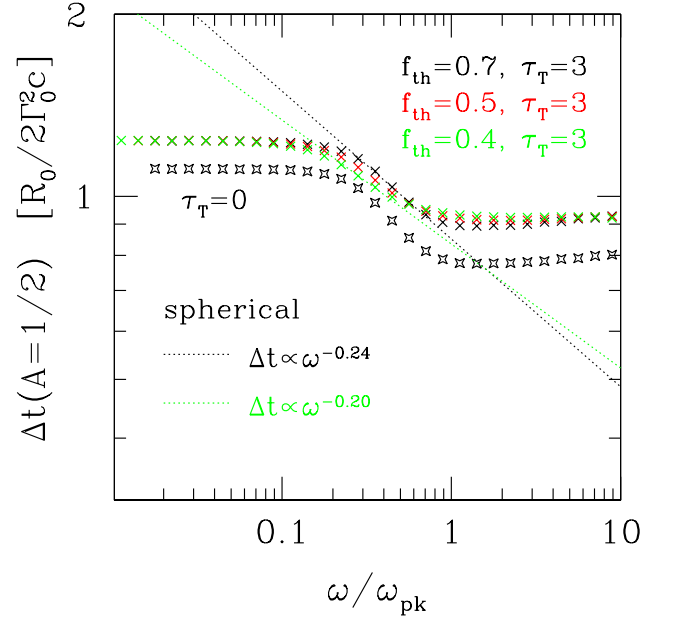}
\caption{Dependence of pulse width on photon energy in the spectrum of Figures \ref{fig:scattspec} and \ref{fig:pulse}
($f{\rm th} = 0.7$, black points).  Pulse width depends more weakly on energy in outflows with higher heating rates and
harder high-energy spectra (red and green points).  Open symbols:  $\tau_T = 0$, $f_{\rm th} = 0.7$.}
\label{fig:pulsewidth}
\vskip .2in
\end{figure}

We evaluate the pulse width $\Delta t(A=1/2)$ by calculating the auto-correlation
of the light curve, as in \cite{fenimore95}, and setting $A = 1/2$.  The energy is divided into bins of width $\Delta\log(\omega) = 0.1$.
The result is shown in Figure \ref{fig:pulsewidth}.  We find, as noted by \cite{qin05}, that the pulse width
narrows over a limited range of energy below the spectral peak, but not above.  The slope of the relation
is a bit steeper for lower heating rates (that is, higher ratios of seed thermal photon energy to that injected in heat,
$f_{\rm th} \equiv \ell_{\rm th}/\ell_{\rm heat}$).  In all cases the relation is a bit shallower than that inferred
from BATSE data, $\Delta t(A=1/2) \sim \omega^{-0.4}$ \citep{fenimore95}.

\begin{figure}
\includegraphics[width=0.48\textwidth]{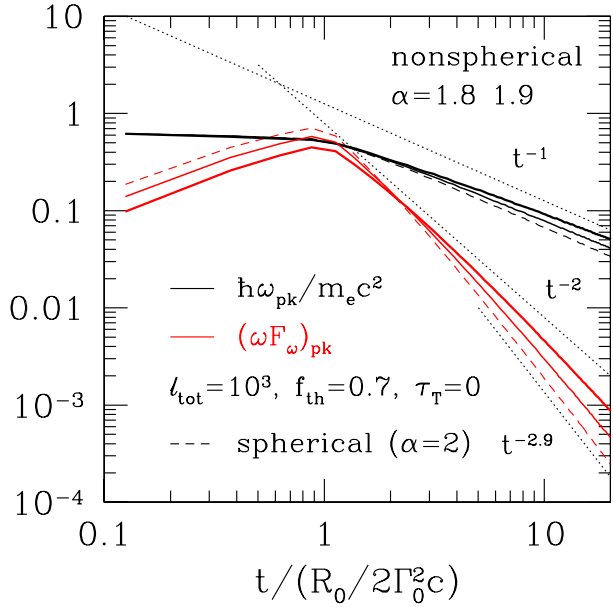}
\caption{Time variation of spectral peak energy and energy flux at the spectral peak, for optically thin emission
from non-spherical shells of a finite emission width $R_0/2\Gamma_0^2 c$.  Thick (thin) solid curves: non-spherical shell 
with curvature delay $t_{\rm curve}(\theta) \propto \theta^{\alpha}$ with $\alpha = 1.8 (1.9)$ and
uniform Lorentz factor, $\Gamma(\theta) = \Gamma_0$.}  
\label{fig:flux_peak_nosph}
\vskip .2in
\end{figure}
\begin{figure}
\includegraphics[width=0.48\textwidth]{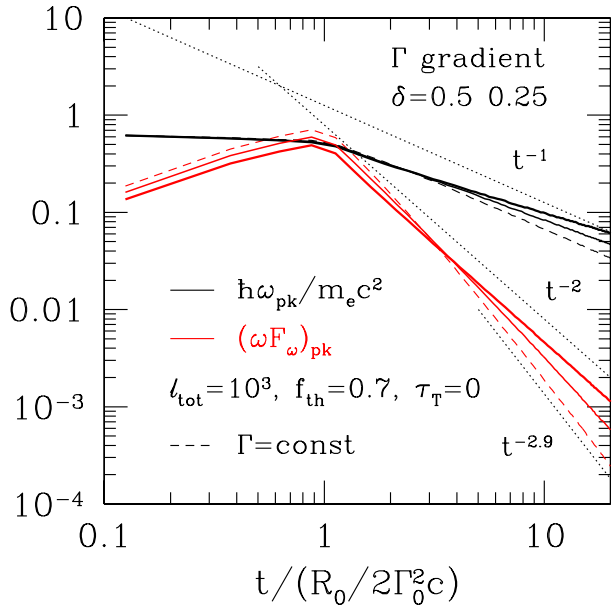}
\caption{Time variation of spectral peak energy and energy flux at the spectral peak, for optically thin emission
from non-spherical shells of a finite emission width $R_0/2\Gamma_0^2 c$.  Thick (thin) solid curves:  spherically curved shell
with latitudinal gradient in Lorentz factor, $\Gamma(\theta) = \Gamma_0(1+\Gamma_0^2\theta^2)^{-\delta/2}$
with $\delta = 0.5 (0.25)$.}
\label{fig:flux_peak_nosph2}
\vskip .2in
\end{figure}
\begin{figure}
\includegraphics[width=0.48\textwidth]{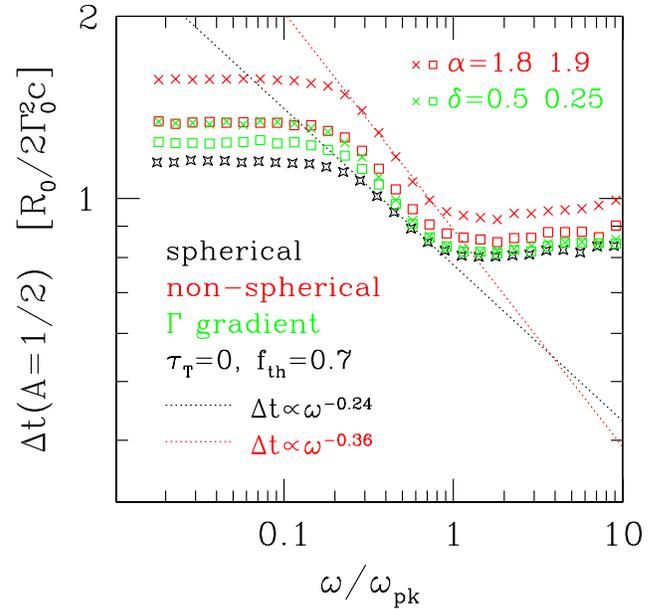}
\caption{Dependence of pulse width on photon energy in the emission geometries of Figure \ref{fig:flux_peak_nosph}.
Red crosses (squares): non-spherical shell 
with curvature delay $t_{\rm curve}(\theta) \propto \theta^{\alpha}$ with $\alpha = 1.8 (1.9)$ and
uniform Lorentz factor, $\Gamma(\theta) = \Gamma_0$.  Green crosses (squares):   spherically curved shell
with latitudinal gradient in Lorentz factor, $\Gamma(\theta) = \Gamma_0(1+\Gamma_0^2\theta^2)^{-\delta/2}$
with $\delta = 0.5 (0.25)$.}
\label{fig:pulsewidth_nosph}
\vskip .2in
\end{figure}
\begin{figure}
\includegraphics[width=0.48\textwidth]{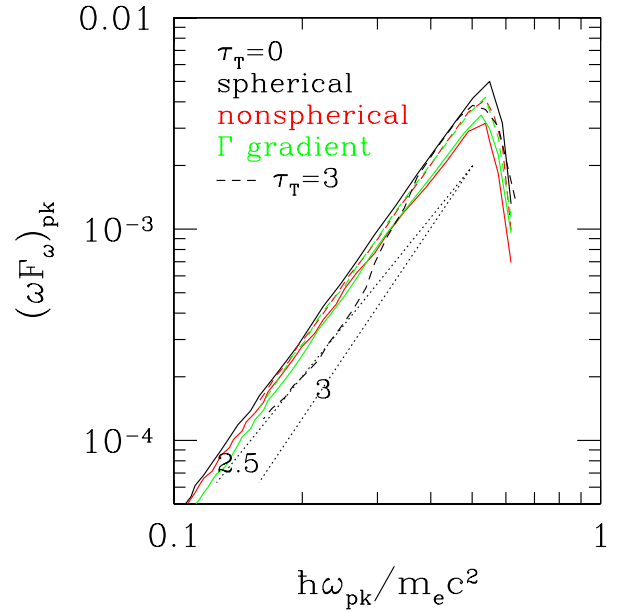}
\caption{Dependence of luminosity at the spectral peak on peak energy, in outflows of various geometries.
Dashed black curve: spherical $\ell_{\rm tot} = 10^3$, $f_{\rm th} = 0.7$ with $\tau_{\rm T} = 3$
at end of heating.  Solid black curve:  same but with $\tau_T$ set to 0.  Solid/dashed red curves show
optically-thin emission from non-spherical shells with uniform $\Gamma$, and solid/dashed green curves show 
optically-thin emission from spherical shells with non-uniform $\Gamma$, in all cases with the
same parameters as in Figure \ref{fig:pulsewidth_nosph}.}
\label{fig:peakflux_nosph}
\vskip .2in
\end{figure}

\subsubsection{Imprint of Non-spherical Shell Structure}\label{s:jetemission}

We now consider an emitting shell which either is i) non-spherically curved; or ii) in which Lorentz factor is a function of 
latitude (angle from the point on the shell that is closest to the observer).  To avoid the complications that arise
from the dynamics of a non-spherical shell, we only consider here an impulsive burst of radiation from an optically
thin shell. 

We first repeat the calculation of pulse decay.
The curvature delay of a non-spherical shell is parameterized as in Section \ref{s:pulsescale}, and
the Lorentz factor profile follows equation (\ref{eq:gamprof}).  We consider four cases:  curvature delay
$t_{\rm curve} \propto \theta^\alpha$ with $\alpha = 1.8, 1.9$ (Figure \ref{fig:flux_peak_nosph}) 
and also the asymptotic scaling $\Gamma(\theta) \propto \theta^{-\delta}$ with $\delta = 0.25, 0.5$ (Figure \ref{fig:flux_peak_nosph2}).  
Next we repeat the calculation of pulse width as a function of photon energy:
the result in Figure \ref{fig:pulsewidth_nosph} shows a somewhat steeper relation
than the spherical calculation in Figure \ref{fig:pulsewidth}, and in closer
agreement with the typical scaling $\Delta t \propto \omega^{-0.4}$
\citep{fenimore95}.  One observes, e.g., that $\alpha = 1.8$ gives pulse flux decay $(\omega F_\omega)_{\rm pk}
\sim t_{\rm obs}^{-2}$ during the early stage of a pulse as well as a frequency scaling
$\Delta t(A=1/2) \propto \omega^{-0.36}$.  However, the decay of $\omega_{\rm pk}$ is now a bit shallower
than $t_{\rm obs}^{-1}$.  

The relation between $(\omega F_\omega)_{\rm pk}$ and $\omega_{\rm pk}$ is shown in
Figure \ref{fig:peakflux_nosph} for optically thin emission from the same types of non-spherical shells.
For reference we also show i) optically-thin emission from a spherical shell; and ii) emission from a spherical
shell with a scattering depth $\tau_T = 3$ in cold pairs.  In the spherical case, the relation of peak flux
to the peak frequency remains closer to $\omega_{\rm pk}^3$ over the full range of the decay; that is, there
is not the same type of softening that one encounters with the time scaling (Figure \ref{fig:flux_peak}).

Non-spherical effects do tend to flatten the $(\omega F_\omega)_{\rm pk} - \omega_{\rm pk}$ relation.  We conclude
that they bring the various measures of pulse decay close to those observed (at least, in the BATSE bands).  
However, introducing a cold scattering photosphere forces a steepening of the
$(\omega F_\omega)_{\rm pk} - \omega_{\rm pk}$ relation beyond the first stages of pulse decay.  
The sign of this effect is easily seen by noting that the photosphere is
expanded for obliquely propagating photons (e.g. \citealt{peer08}), so a given $\omega_{\rm pk}$ corresponds to
a later time and a lower flux.   

One degree of freedom that we have not considered here, but which plausibly
is present during breakout of the magnetic field, is an intrinsic latitudinal
gradient of $\omega_{\rm pk}$.  This may need to be invoked
to obtain consistency with the observed scaling between peak flux and
spectral peak energy.

\subsection{Outflow Heated Continuously from a Large Scattering Depth}

A relativisic outflow that is heated continuously outward from a large scattering
depth develops an extended, high-energy spectral tail to a seed thermal radiation field
\citep{giannios06b,beloborodov10}.  Here we examine how the pulses of emergent radiation
depend on photon energy.  There is a much stronger effect as compared with a cold scattering screen (Sections \ref{s:coldscatt} and \ref{s:jetemission}).

Heating is assumed to continue from inside to outside the photosphere, as in the most recent
calculation of \cite{giannios08}.  
The outflow starts at a certain initial scattering depth $\tau_{\rm T}(R_0)$
evaluated at the radius $R_0$ where heating begins.  Beyond this point, the comoving 
particle energy $\gamma_e m_ec^2$ adjusts so that
\be\label{eq:yC}
{4\over 3}\left(\langle \gamma_e^2\rangle -1\right) n_e \sigma_T {r\over 2\Gamma^2} = {dy_{\rm C}\over d\ln t} = {\rm const}.
\ee
We choose a constant Lorentz factor $\Gamma$ and spherical geometry, 
and neglect any effect of pair creation or annihilation.
Then $n_e(r) \propto r^{-2}$ since $\Gamma \gg 1$.
\begin{figure}
\includegraphics[width=0.48\textwidth]{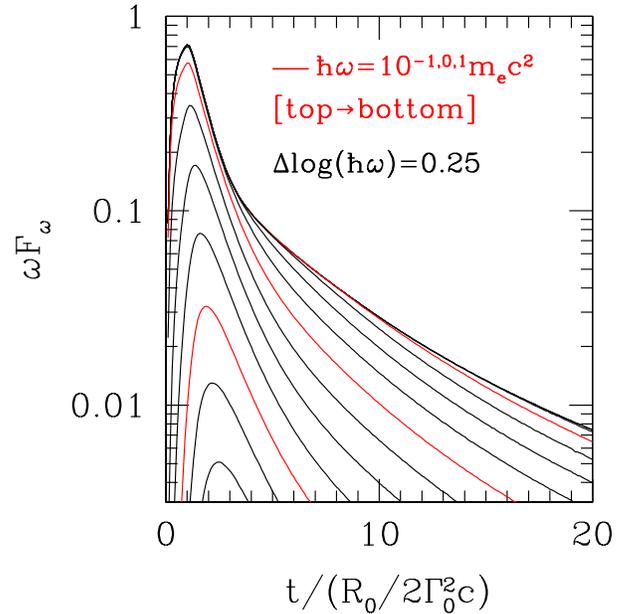}
\caption{Pulse width as a function of photon energy for a multiple scattering photosphere with
 $\tau_{T,0} = 3$ at base of heating layer, and $dy_{\rm C}/d\ln t = 1.5$.} 
\label{fig:pulse_mult}
\vskip .2in
\end{figure}
\begin{figure}
\includegraphics[width=0.48\textwidth]{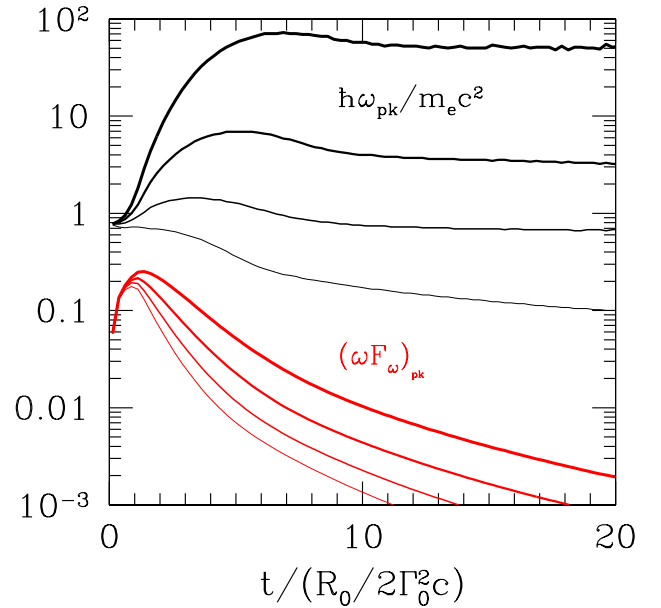}
\caption{Variation of peak energy and energy flux at the spectral peak, as a function of time.
Spherical, relativistic outflow is continuously heated from inside its photosphere.
Scattering depth $\tau_{T,0} = 3$ at base of heating layer, and $dy_{\rm C}/d\ln t = 2$, 1.5, 1, 0.5 (thick to thin lines).}
\label{fig:pulse_train_mult}
\vskip .2in
\end{figure}
\begin{figure}
\includegraphics[width=0.48\textwidth]{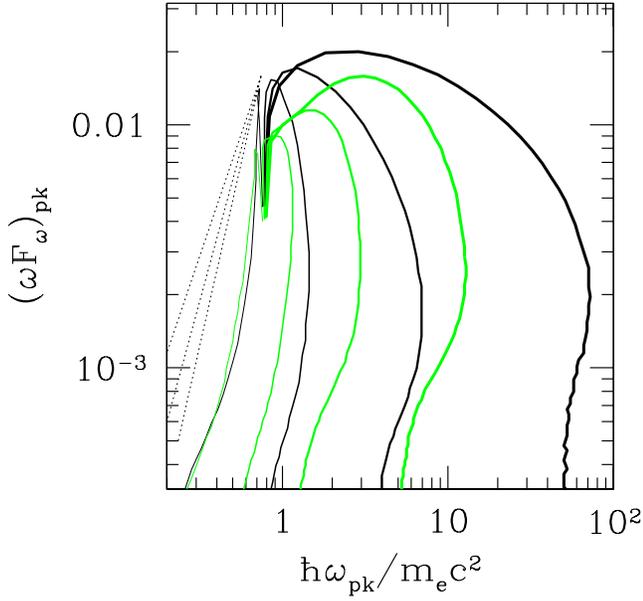}
\caption{Dependence of energy flux at the spectral peak on peak energy in a spherical, 
relativistic outflow that is continously heated from inside its photosphere.
Green (black) lines:  scattering depth $\tau_{T,0}=3$ (10) at base of heating layer, and 
$dy_{\rm C}/d\ln t = 2$, 1.5, 1, 0.5 (thick to thin).}
\label{fig:peak_mult}
\vskip .2in
\end{figure}
\begin{figure}
\includegraphics[width=0.48\textwidth]{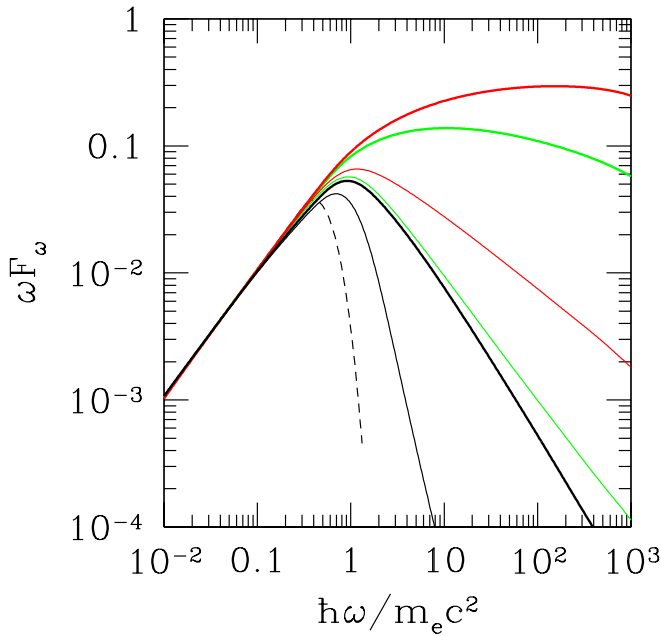}
\caption{Effect of widening the shell on the output spectrum produced by continuous heating.
Heating starts at radius $R_0$ and scattering depth $\tau_{T,0} = 3$, and Lorentz factor constant
$\Gamma = \Gamma_0 = 10^2$.  Shell width in units of $R_0/2\Gamma_0$:  1, 10, 30 corresponds
to black, green and red lines.  $dy_{\rm C}/d\ln t = 1$ (0.5) corresponds to heavy (light) lines.}
\label{fig:spectrum_mult_long}
\vskip .2in
\end{figure}
\begin{figure}
\includegraphics[width=0.48\textwidth]{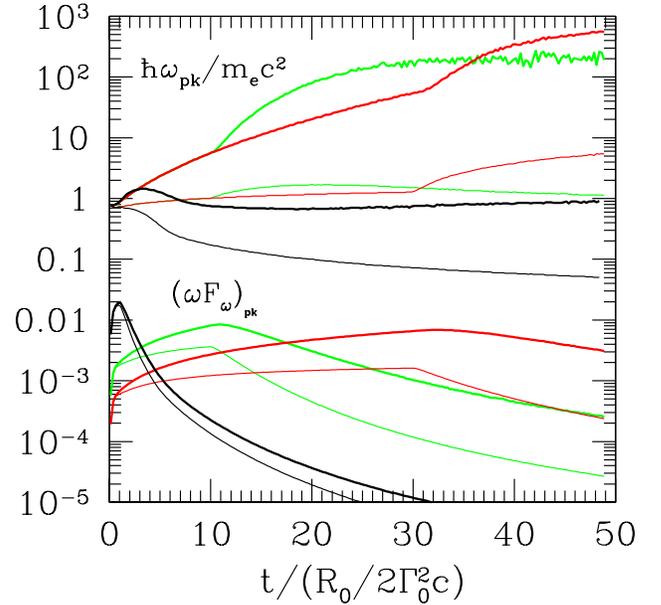}
\caption{Effect of widening the shell on the evolution of spectral hardness and flux at the
spectral peak, in the case of continuous heating.  Same models as Figure (\ref{fig:spectrum_mult_long}.
Shell width in units of $R_0/2\Gamma_0$:  1, 10, 30 corresponds
to black, green and red lines.  $dy_{\rm C}/d\ln t = 1$ (0.5) corresponds to heavy (light) lines.}
\label{fig:peak_mult_long}
\vskip .2in
\end{figure}

The output spectrum, as shown in Figure 17 in Paper II, confirms the formation of extended high-energy
tail, with an increasing hardness as $y_{\rm C}$ is raised.  Since harder photons are created by multiple
scattering of softer photons, the pulses are broader at high energies (Figure \ref{fig:pulse_mult}), in contradiction
with observations:  indeed the burst shows strong soft-to-hard evolution, again in contradiction with
most GRB behavior (Figure \ref{fig:pulse_train_mult}).    Another difficulty with the model is revealed
by correlating the spectral peak energy with the energy flux at the peak (Figure \ref{fig:peak_mult}).

The preceding calculations started with an emitting shell that is causally connected in the radial direction,
$\Delta R_{\rm em} = R_0/2\Gamma_0^2$.  To see that the soft-to-hard evolution is not an artifact of this
assumption, we considers wider shells that are 10-30 times wider.  The resulting spectrum is shown in
Figure \ref{fig:spectrum_mult_long} and the evolution of the spectral peak in Figure \ref{fig:peak_mult_long}.
In this case also, the time evolution of a pulse is inconsistent with the behavior of GRBs.  

\section{Discussion}\label{s:discuss}

We have investigated the pulse structure resulting from the breakout of an ultraluminous, magnetized outflow from a 
confining baryonic medium, followed by a delayed reheating at low scattering depth.   We have also
considered the imprint of multiple scattering on radiation pulses emitted by curved, relativistic shells.  
Some of our key conclusions are summarized as follows.

{\it Pulse Distribution and Null Periods in GRBs.}  We hypothesize that the period of engine
activity in GRBs is typically shorter than the observed $T_{90}$ duration.  In long GRBs, it is better characterized by 
$\Delta t \sim 1$ peak of the pulse width distribution \citep{norris96} and power-spectrum break \citep{bss00}.  These measurements
are consistent with the $0.5$-$3$ s timescale for the collapse of the central cores of Wolf-Rayet stars.  
Baryonic material that is swept up by a shell
of relativistic magnetofluid will corrugate at a radius $R_{\rm br} > 2\Gamma_{\rm br}^2 c\Delta t$, allowing the
magnetofluid to break through in independent streams that then freely expand to a higher Lorentz factor.  These streams
emit separate gamma-ray pulses that need not overlap in time.  

{\it X-ray Flares.}  We apply the same model of magnetic breakout from a baryon shell to the X-ray flares that are 
observed in the tails of GRBs \citep{chincarini10,margutti10}, the main difference being an increased baryon 
loading relative to the energy carried by the entrained magnetic field.  This allows the delayed release of energy 
from material that was energized at the same time as the faster, relativistic fluid that emits the prompt gamma-ray pulse.  
The need for delayed accretion onto the engine is therefore called into question.

{\it Hard-to-soft Evolution.}  Emission from a curved spherical shell generally shows hard-to-soft spectral evolution,
with pulses peaking earlier at higher energies \citep{shenoy13}.  This 
trend is a key observed feature of GRB pulses, and holds even for quasi-thermal radiation that is lacking a
high-energy power-law tail.  We show that the trend is preserved by pair breakdown at a low scattering depth, which we
described in Paper II as a mechanism for generating a high-energy tail.

{\it Inconsistency between an extended, Comptonizing photosphere and observed GRB pulse pehavior.}  An alternative
hypothesis is that the high-energy spectrum of a GRB forms in close analogy with an accretion disk corona, by
diffusive upscattering of softer thermal photons \citep{giannios06b,lazzati10}.  We confirm that a high-energy spectral tail can form by such
a mechanism, but point out two disagreements with observation:  the spectral peak tends to become very broad
if the high-energy tail is hard;  and, more seriously, the harder photons tend to lag softer ones.  We conclude
that multiple scattering at a photosphere in a relativistic outflow is not a viable explanation for high-energy
spectral tails in GRBs.  Residual scattering following impulse heating
and rapid pair breakdown is found to have a much milder effect on pulse widths. 

{\it Outflow Lorentz Factor at Breakout.}  The escape of magnetofluid from a forward baryon shell is a casual process,
occuring on a angular scale $\delta\theta \lesssim 1/\Gamma_{\rm br}$.  
It should also be noted that off-axis emission is still expected here:  the frozen pairs that are advected beyond the
breakout point reach $\tau_{\rm T}^\pm \sim 1$ only after the Lorentz factor has increased by a factor $\gtrsim 3$ beyond the
breakout value.  

{\it Imprint of shell shape and angular gradients in Lorentz factor on pulse behavior.}   The typical GRB pulse cooling
behavior, as represented by the scalings $(\omega F_\omega)_{\rm pk} \sim t_{\rm obs}^{-2}$ and $(\omega F_\omega)_{\rm pk} \sim 
\omega_{\rm pk}^{2-2.5}$ \citep{fenimore95,borgonovo01,ryde02,ghirlanda10,preece14} are in disagreement with impulsive emission
from a spherical, relativistic shell.  We have shown that the initial decay is flatter than $t_{\rm obs}^{-3}$ when 
the effect of finite shell width is taken into account.  

We have also explored the effect of introducing non-spherical shell curvature, as well
as a downward gradient in Lorentz factor away from the point on the shell that is closest to the observer.  Analytic scalings
are derived in the optically thin and thin-shell regime.  It is found that a flatter scaling of $(\omega F_\omega)_{\rm pk}$
with both time and peak frequency is obtained when the emitting shell is flatter than a sphere (curvature delay
increasing more slowly than $\sim \theta^2$) and when $\Gamma(\theta)$ decreases with latitude.

In addition, we have explored the effect of finite shell width and
of photospheric scattering on pulse decay, using a Monte Carlo method.  
Both effects tend to flatten the decay of pulses.   The flattening effect of 
non-spherical curvature and $d\Gamma/d\theta < 0$ is also reproduced.  
However, consistency with the observed scaling of 
$(\omega F_\omega)_{\rm pk}$ with $\omega_{\rm pk}$ may require the presence 
of a latitudinal temperature gradient at the breakout of the magnetic 
field, an effect that is not included in our calculations.
The combined effect of scattering and non-spherical flow divergence 
has yet to be explored, and deserves further study.

\subsection{Some Outstanding Issues}

{\it Plateau in the $T_{90}$ Distribution of GRBs.}  \cite{bromberg12} find a flattening of the duration 
distribution within a subset of GRBs,\footnote{Spectrally soft BATSE bursts and Swift bursts.} 
over the interval $1 \lesssim T_{90} \lesssim 10$ s.  They suggest that this flattening is a signature of
jet breakout from a Wolf-Rayet star,  working from the usual assumption that the upper envelope of $T_{90}$ measures 
the main active period of the engine.  They note that in some bursts most of the jet energy is swallowed before the 
breakout of the jet head (see equation (\ref{eq:ecocoon})).  If the energy that is
absorbed by the relativistic jet cocoon is {\it not} a potential source 
of GRB emission (the opposite of the conclusion reached in Section \ref{s:tbreak}), then one expects to see a 
number of bursts whose $T_{90}$ is significantly {\it shorter} than the engine activity.  

We note that the lower end of the plateau approximately coincides with the peak of the pulse distribution in long GRBs.  
This leads, following the discussion in Sections \ref{s:null}, \ref{s:tbreak}, to an alternative interpretation of the plateau:
that it represents a {\it lengthening} of $T_{90}$ due to delayed breakout of a magnetic field from a thin baryon shell.
The longer that $T_{90}$ is compared with $\sim 1$ s, the larger the curvature delay across the visible part of the shell.
Further progress might be made in clarifying this issue by considering the pulse structure of the bursts in the plateau zone
of $T_{90}$.  

{\it Narrower pulses above the gamma-ray spectral peak.}  
Coincident X-ray and gamma-ray measurements of the GRB prompt emission show a strong broadening of the 
pulses at energies $\sim 10^{-2} \hbar\omega_{\rm pk}$ \citep{butler07,preece14}.  The trend in the bands probed by
BATSE is typically $\Delta t \propto \omega^{-0.4}$ \citep{fenimore95}.  
Most BATSE GRBs show the narrowest pulses in the highest-energy band, which usually lies above $\omega_{\rm pk}$ (e.g. \citealt{peng12}, 
and references therein).  Fermi measurements have probed an expanded range of energies above the peak 
(e.g. \citealt{preece14}), but the scaling above the peak is not yet well constrained.  

Energy-dependent cooling is an obvious candidate for explaining this broad trend of pulse width with energy.   The difficulty
here is that GRB emission occurs at a high compactness unless the outflow Lorentz factor is extremely high, so
that synchrotron cooling should be rapid at all frequencies.  

It has been understood for some time that relativistic aberration leads to a widening of pulses below the peak, even if the emission time is 
independent of photon energy, but only over a restricted range of energies:  typically between $\sim 0.2\omega_{\rm pk}$ and
$\omega_{\rm pk}$ \citep{qin05}.  This effect does not extend above the peak and, in addition, cannot explain a strong
broadening in the X-ray band.

The high-energy spectral tail forms at a much higher Lorentz factor in the model examined here and in Paper II.  
As a result, the high-energy flux can divide into narrower pulses above the 
peak (see equation (\ref{eq:tvary})).
For this reason, measurements of pulse width near and above the spectral peak 
provide valuable diagnostics of the
emission mechanism:  in particular, of a division of the emission into two steps, an early thermalization phase followed by reheating.

{\it Strong pulse broadening in the X-ray band.}
The fact remains that the shell curvature effect should be present.
Then the strong broadening of pulses in the X-ray band may involve additional emission.  For example, the width of the well-measured first pulse of GRB 130427A
varies only slowly over a decade of photon energy below the peak, but a stronger broadening is found in the X-ray band \citep{preece14}.

We return to the magnetic breakout model developed in Sections \ref{s:null} and \ref{s:baryon}.  Here the baryonic component
of the shell is a source of softer photons \citep{thompson06}.  We have argued that entrainment of baryons by the relativistic (magnetized) outflow is 
achieved after they become thin enough that the black body radiation within them is forced out by the applied magnetic pressure.   
The effective temperature of this radiation is much lower than than the spectral peak of the frozen pair plasma in the magnetic
wind (as it would also be in more standard models of a jet cocoon, e.g. \citealt{rr02}).  

Setting ${1\over 3}a {T_{\rm bb}'}^4$
equal to the comoving magnetic pressure, with total magnetic energy $E_{\rm P}$ in a shell of width $c\Delta t$ and radius $R_{\rm br}$
given by equation (\ref{eq:rbreak}), one easily finds an observed temperature
\be
{4\Gamma_{\rm br}\over 3}\kB T_{\rm bb}' = {1.3\,{\rm keV}\over
(\Delta t/3~{\rm s})^{3/4}} {E_{\rm P,52}^{1/4}\over ({\cal R}_{\rm br}/10)^{3/4}}
\left({\Gamma_{\rm br}\over 3}\right)^{-1/2}.
\ee
Further changes in radiation temperature due to adiabatic losses and/or reheating of the baryons are obviously possible.

{\it Anisotropic emission.}  Although anisotropic heating of embedded $e^\pm$ along the background magnetic field
plays an important role in regulating the relative amplitudes of Compton and synchrotron emission (\citealt{thompson06},
paper II), we do not find a compelling need for anisotropic emission as a source of pulse variability.  Nonetheless,
it may still play some role in forming pulse sub-structure at high energies, especially in sub-pulses that do not show
any evidence for off-axis emission in a decaying tail.

A compelling need for magnetic reconnection is also lacking.   Reconnection freezes out after the breakout of 
magnetofluid from a forward baryonic layer,
due to the contraction in the radial causal horizon in the outflow.  Although a magnetized jet can still be expected
to contain a radial current sheet structure, as driven by stochasticity in a dynamo process operating in the engine, the 
resulting flips of the non-radial magnetic field may be quite widely spaced.  Localized bursts of bulk relativistic 
motion, driven by magnetic reconnection (e.g. \citealt{lyutikov03}), are therefore disfavored as a dominant source of
pulse variability.  More detailed discussion of reconnection can be found in Papers I and II.

{\it Pulse structure above $\sim 1$ GeV.}  There is a systematic reduction in the number of pulses in GRBs
detected by Fermi at high photon energies \citep{ackermann13}.  Many bursts show one pulse, consistent with the
a delayed onset of high-energy emission at a single forward shock.  However, some bursts show more than
one pulse.  This remains consistent with emission from the forward shock if breakout of the relativistic flow occurs
in causally separated patches, which remain disconnected in the angular direction but have overlapping Lorentz
cones (see Figure \ref{fig:breakout} and the discussion in Sections \ref{s:clues}, \ref{s:baryon}).

{\it Side-scattering of the prompt emission during pair-loading of an external wind.}  The medium outside
the forward shock is loaded with electron-positron pairs, after which it is forced outward to relativistic
speeds by the intense gamma-ray flux \citep{tm00,beloborodov02}.  This process of pre-acceleration generally
shuts off as the relativistic ejecta are coming into contact with the forward shock \citep{thompson06}.  
As a result, a signficant fraction of the prompt gamma-rays begin to stream across the forward shock while
the external medium is still pair loaded, but not sufficiently to develop relativistic motion (pair multiplicity
${\cal M}_\pm = 2n_{e^+}/n_p \lesssim 10^2$.  

Consider, for example, an external Wolf-Rayet wind of mass loss rate $\dot M_w = 10^{-5}\,M_\odot$ yr$^{-1}$ 
in which this critical pair loading is reached at $r \sim 3\times 10^{15}$ cm.  Then the scattering depth of 
the wind material, moving at $V_w \sim 10^3$ km s$^{-1}$, is $\tau_{\rm T} = (1+{\cal M}_\pm) 
\kappa_{\rm es}\dot M_w/4\pi V_w r \sim 0.02$.   Photons side-scattered in this way will appear as a broad
tail extending across a significant part of the burst.  The effect is stronger in GRBs of a lower isotropic energy
and shorter duration, in which pair loading shuts off at a smaller radius.


\begin{appendix}

\section{A. Monte Carlo Evaluation of Output Photon Spectrum and Pulse Profiles}\label{s:monte}

We follow the trajectories of individual photons in the rest frame of the engine, and calculate their scattering
in the Thomson approximation.   The photons are released within a spherical shell of radial thickness
$\Delta R_{\rm em}$ at radius $\leq R_0$.  The scattering charges are assumed to move relativistically, and the photon trajectory
and Doppler boosts to and from the comoving frame are worked out in the small-angle approximation.
The observed time of each photon is recorded as described in Section \ref{s:scatter}.

The position $\xi_{\rm em}$ of emission within the shell is drawn randomly, and the 
photons are emitted isotropically in a frame moving at Lorentz factor (\ref{eq:gamprof}).   The case of spherical
symmetric expansion ($\delta_\theta = 0$) is simpler:  in this case, we need only keep track of the angle $\psi
= \cos^{-1}(\hat k\cdot \hat r)$ between the propagation direction $\hat k$ of a photon and the radial flow direction.
In a more general jet geometry, we
keep track of the direction cosines of each photon, as defined with respect to the symmetry axis of the jet, using
Cartesian coordinates.  The only subtlety here involves the transformation of the photon wavevector back to the 
engine frame after scattering, and the re-evaluation of the direction cosines.

The electron density is evolved passively outward, $n_e \propto r^{-2}$ given the ultra-relativistic motion, representing
the freeze-out of pairs following a dissipation episode.  
When $\Gamma$ does depend on $\theta$, we maintain a uniform scattering coefficient across $\theta$, so that
$n_e(\theta,r)/\Gamma^2(\theta,r)$ depends only on radius.  This is motivated by the rapid evolution of the 
scattering depth in pairs to a characteristic value $\tau_{\rm T} \sim 3$ at the end of heating and onset of free expansion
\citep{tg13}.  More general angular profiles can be considered, but this simplification allows us to focus on the consequences
of introducing an angular gradient to $\Gamma$.   If this gradient is negative ($\delta_\theta > 0$)
and the opacity is not buffered in the way just described, then broad scattering tails appear in pulse profiles, which
are not observed in GRBs.  In cases where the flow is aspherical, we start photons over an angular range $\sim 10/\Gamma_0$ 
and record them at a fixed observer direction.  

Scattering in a cold shell is performed using the Thomson angular distribution for the outgoing photon.
After transforming $\psi$ to the local comoving frame, we pick scattering angles $\theta_s'$,
$\phi_s'$ with respect to the flow direction.  The direction cosine of the outgoing photon is determined via 
$\mu_{\rm em}' = \mu'\cos\theta_s' + (1-\mu'^2)^{1/2}\sin\theta_s'\cos\phi_s'$, followed by a boost to the stellar frame.
We have checked that the assumption of isotropic scattering produces spectra and pulse profiles that do not differ
measurably from this exact treatment.  When the scattering particles are heated, that allows us to adopt 
an isotropic source function (which is an exact property of any relativistic and locally isotropic particle distribution).  
Then the energy shift in the comoving frame is decoupled from the boost back to the rest frame of the engine.  
The temperature of the pairs in a continuously heated shell is adjusted using equation (\ref{eq:yC}).

\end{appendix}



\begin{thebibliography}{}







\bibitem[Ackermann et al.(2013)]{ackermann13} Ackermann, M., Ajello, M., Asano, K., et al.\ 2013, \apjs, 209, 11 
\bibitem[Ackermann et al.(2014)]{ackermann14} Ackermann, M., Ajello, M., Asano, K., et al.\ 2014, Science, 343, 42 
\bibitem[Amati et al.(2002)]{amati02} Amati, L., Frontera, F., Tavani, M., et al.\ 2002, \aap, 390, 81 
\bibitem[Asano \& M{\'e}sz{\'a}ros(2011)]{asano11} Asano, K., \& M{\'e}sz{\'a}ros, P.\ 2011, \apj, 739, 103 
\bibitem[Asano \& M{\'e}sz{\'a}ros(2013)]{asano13} Asano, K., \& M{\'e}sz{\'a}ros, P.\ 2013, \jcap, 9, 8 
\bibitem[Begelman \& Cioffi(1989)]{bc89} Begelman, M.~C., \& Cioffi, D.~F.\ 1989, \apjl, 345, L21 
\bibitem[Beloborodov et al.(2000)]{bss00} Beloborodov, A.~M., Stern, B.~E., \& Svensson, R.\ 2000, \apj, 535, 158 
\bibitem[Beloborodov(2002)]{beloborodov02} Beloborodov, A.~M.\ 2002, \apj, 565, 808
\bibitem[Beloborodov(2010)]{beloborodov10} Beloborodov, A.~M.\ 2010, \mnras, 407, 1033 
\bibitem[Beloborodov et al.(2013)]{beloborodov13b} Beloborodov, A.~M., Hascoet, R., \& Vurm, I.\ 2013, arXiv:1307.2663 
\bibitem[Blandford \& Znajek(1977)]{bz77} Blandford, R.~D., \& Znajek, R.~L.\ 1977, \mnras, 179, 433 
\bibitem[Borgonovo \& Ryde(2001)]{borgonovo01} Borgonovo, L., \& Ryde, F.\ 2001, \apj, 548, 770 
\bibitem[Broderick(2005)]{broderick05} Broderick, A.~E.\ 2005, \mnras, 361, 955 
\bibitem[Bromberg et al.(2012)]{bromberg12} Bromberg, O., Nakar, E., Piran, T., \& Sari, R.\ 2012, \apj, 749, 110 
\bibitem[Bromberg et al.(2014)]{bromberg14} Bromberg, O., Granot, J., Lyubarsky, Y., \& Piran, T.\ 2014, arXiv:1402.4142 
\bibitem[Butler \& Kocevski(2007)]{butler07} Butler, N.~R., \& Kocevski, D.\ 2007, \apj, 663, 407 
\bibitem[Chincarini et al.(2010)]{chincarini10} Chincarini, G., Mao, J., Margutti, R., et al.\ 2010, \mnras, 406, 2113 
\bibitem[Daigne \& Mochkovitch(1998)]{daigne98} Daigne, F., \& Mochkovitch, R.\ 1998, \mnras, 296, 275 
\bibitem[Dermer(2004)]{dermer04} Dermer, C.~D.\ 2004, \apj, 614, 284 
\bibitem[Dessart et al.(2009)]{dessart09} Dessart, L., Ott, C.~D., Burrows, A., Rosswog, S., \& Livne, E.\ 2009, \apj, 690, 1681 
\bibitem[Drenkhahn \& Spruit(2002)]{drenkhahn02} Drenkhahn, G., \& Spruit, H.~C.\ 2002, \aap, 391, 1141 
\bibitem[Duffell \& MacFadyen(2014)]{duffell14} Duffell, P., \& MacFadyen, A.\ 2014, arXiv:1403.6895 
\bibitem[Eichler \& Levinson(2000)]{eichler00} Eichler, D., \& Levinson, A.\ 2000, \apj, 529, 146 
\bibitem[Eichler(2014)]{eichler14} Eichler, D.\ 2014, \apjl, 787, L32 
\bibitem[Falcone et al.(2006)]{falcone06} Falcone, A.~D., Burrows, D.~N., Lazzati, D., et al.\ 2006, \apj, 641, 1010 
\bibitem[Falcone et al.(2007)]{falcone07} Falcone, A.~D., Morris, D., Racusin, J., et al.\ 2007, \apj, 671, 1921 
\bibitem[Fenimore et al.(1995)]{fenimore95} Fenimore, E.~E., in 't Zand, J.~J.~M., Norris, J.~P., Bonnell, J.~T., 
\& Nemiroff, R.~J.\ 1995, \apjl, 448, L101 
\bibitem[Fenimore et al.(1996)]{fenimore96} Fenimore, E.~E., Madras, C.~D., \& Nayakshin, S.\ 1996, \apj, 473, 998 
\bibitem[Ghirlanda et al.(2010)]{ghirlanda10} Ghirlanda, G., Nava, L., \& Ghisellini, G.\ 2010, \aap, 511, A43 
\bibitem[Giannios(2006)]{giannios06a} Giannios, D.\ 2006, \aap, 455, L5 
\bibitem[Giannios(2006)]{giannios06b} Giannios, D.\ 2006, \aap, 457, 763 
\bibitem[Giannios(2008)]{giannios08} Giannios, D.\ 2008, \aap, 480, 305 
\bibitem[Giannios \& Spruit(2006)]{giannios06c} Giannios, D., \& Spruit, H.~C.\ 2006, \aap, 450, 887 
\bibitem[Gill \& Thompson(2014)]{gt14} Gill, R., \& Thompson, C., 2014, \apj, submitted
\bibitem[Heinz \& Begelman(1999)]{heinz99} Heinz, S., \& Begelman, M.~C.\ 1999, \apjl, 527, L35 
\bibitem[Jiang et al.(2013)]{jiang13} Jiang, Y.-F., Davis, S.~W., \& Stone, J.~M.\ 2013, \apj, 763, 102 
\bibitem[Kobayashi et al.(1997)]{kobayashi97} Kobayashi, S., Piran, T., \& Sari, R.\ 1997, \apj, 490, 92 
\bibitem[Kumar \& Panaitescu(2000)]{kumar00} Kumar, P., \& Panaitescu, A.\ 2000, \apjl, 541, L51 
\bibitem[Kumar \& Piran(2000)]{kumar_piran00} Kumar, P., \& Piran, T.\ 2000, \apj, 535, 152 
\bibitem[Lazar et al.(2009)]{lazar09} Lazar, A., Nakar, E., \& Piran, T.\ 2009, \apjl, 695, L10
\bibitem[Lazzati \& Perna(2007)]{lazzati07} Lazzati, D., \& Perna, R.\ 2007, \mnras, 375, L46 
\bibitem[Lazzati et al.(2009)]{lazzati09} Lazzati, D., Morsony, B.~J., \& Begelman, M.~C.\ 2009, \apjl, 700, L47 
\bibitem[Lazzati \& Begelman(2010)]{lazzati10} Lazzati, D., \& Begelman, M.~C.\ 2010, \apj, 725, 1137 
\bibitem[Lazzati et al.(2013)]{lmmb13} Lazzati, D., Morsony, B.~J., Margutti, R., \& Begelman, M.~C.\ 2013, \apj, 765, 103 
\bibitem[Levinson \& Begelman(2013)]{levinson13} Levinson, A., \& Begelman, M.~C.\ 2013, \apj, 764, 148 
\bibitem[Lyutikov \& Blandford(2003)]{lyutikov03} Lyutikov, M., \& Blandford, R.\ 2003, arXiv:astro-ph/0312347 
\bibitem[MacFadyen \& Woosley(1999)]{macfadyen99} MacFadyen, A.~I., \& Woosley, S.~E.\ 1999, \apj, 524, 262 
\bibitem[Margutti et al.(2010)]{margutti10} Margutti, R., Guidorzi, C., Chincarini, G., et al.\ 2010, \mnras, 406, 2149 
\bibitem[Margutti et al.(2011)]{margutti11} Margutti, R., 
Chincarini, G., Granot, J., et al.\ 2011, \mnras, 417, 2144 
\bibitem[Maselli et al.(2014)]{maselli14} Maselli, A., Melandri, A., Nava, L., et al.\ 2014, Science, 343, 48 
\bibitem[Matzner(2003)]{matzner03} Matzner, C.~D.\ 2003, \mnras, 345, 575 
\bibitem[Meszaros \& Rees(1997)]{meszaros97} Meszaros, P., \& Rees, M.~J.\ 1997, \apjl, 482, L29 
\bibitem[McKinney(2006)]{mckinney06} McKinney, J.~C.\ 2006, \mnras, 368, L30 
\bibitem[McKinney \& Uzdensky(2012)]{mckinney12} McKinney, J.~C., \& Uzdensky, D.~A.\ 2012, \mnras, 419, 573 
\bibitem[Nakar \& Piran(2002)]{nakar02} Nakar, E., \& Piran, T.\ 2002, \mnras, 331, 40 
\bibitem[Narayan \& Kumar(2009)]{narayan09} Narayan, R., \& Kumar, P.\ 2009, \mnras, 394, L117 
\bibitem[Norris et al.(1996)]{norris96} Norris, J.~P., Nemiroff, R.~J., Bonnell, J.~T., et al.\ 1996, \apj, 459, 393 
\bibitem[Paczynski(1998)]{paczynski98} Paczynski, B.\ 1998, \apjl, 494, L45 
\bibitem[Paxton et al.(2013)]{paxton13} Paxton, B., Cantiello, M., Arras, P., et al.\ 2013, \apjs, 208, 4 
\bibitem[Pe'er(2008)]{peer08} Pe'er, A.\ 2008, \apj, 682, 463 
\bibitem[Pendleton et al.(1997)]{pendleton97} Pendleton, G.~N., Paciesas, W.~S., Briggs, M.~S., et al.\ 1997, \apj, 489, 175 
\bibitem[Pe'er et al.(2006)]{peer06} Pe'er, A., M{\'e}sz{\'a}ros, P., \& Rees, M.~J.\ 2006, \apj, 642, 995 
\bibitem[Peng et al.(2012)]{peng12} Peng, Z.~Y., Zhao, X.~H., Yin, Y., Bao, Y.~Y., \& Ma, L.\ 2012, \apj, 752, 132 
\bibitem[Preece et al.(2014)]{preece14} Preece, R., Burgess, J.~M., von Kienlin, A., et al.\ 2014, Science, 343, 51 
\bibitem[Qin et al.(2005)]{qin05} Qin, Y.-P., Dong, Y.-M., Lu, R.-J., Zhang, B.-B., \& Jia, L.-W.\ 2005, \apj, 632, 1008 
\bibitem[Ramirez-Ruiz et al.(2002)]{rr02} Ramirez-Ruiz, E., Celotti, A., \& Rees, M.~J.\ 2002, \mnras, 337, 1349 
\bibitem[Rees \& Meszaros(1994)]{rm94} Rees, M.~J., \& Meszaros, P.\ 1994, \apjl, 430, L93 
\bibitem[Russo \& Thompson(2013a)]{russo13a} Russo, M., \& Thompson, C.\ 2013, \apj, 767, 142 
\bibitem[Russo \& Thompson(2013b)]{russo13b} Russo, M., \& Thompson, C.\ 2013, \apj, 773, 99
\bibitem[Ryde \& Petrosian(2002)]{ryde02} Ryde, F., \& Petrosian, V.\ 2002, \apj, 578, 290 
\bibitem[S{\c a}dowski et al.(2013)]{sadowski13} S{\c a}dowski, A., Narayan, R., Penna, R., \& Zhu, Y.\ 2013, \mnras, 436, 3856 
\bibitem[Sari \& Piran(1997)]{sari97} Sari, R., \& Piran, T.\ 1997, \apj, 485, 270 
\bibitem[Shenoy et al.(2013)]{shenoy13} Shenoy, A., Sonbas, E., Dermer, C., et al.\ 2013, \apj, 778, 3 
\bibitem[Spruit et al.(2001)]{spruit01} Spruit, H.~C., Daigne, F., \& Drenkhahn, G.\ 2001, \aap, 369, 694 
\bibitem[Swenson \& Roming(2014)]{swenson14} Swenson, C.~A., \& Roming, P.~W.~A.\ 2014, \apj, 788, 30 
\bibitem[Tan et al.(2001)]{tan01} Tan, J.~C., Matzner, C.~D., \& McKee, C.~F.\ 2001, \apj, 551, 946 
\bibitem[Tchekhovskoy et al.(2010)]{tchek10} Tchekhovskoy, A., Narayan, R., \& McKinney, J.~C.\ 2010, 711, 50
\bibitem[Thompson(1994)]{thompson94} Thompson, C.\ 1994, \mnras, 270, 480 
\bibitem[Thompson(2006)]{thompson06} Thompson, C.\ 2006, \apj, 651, 333 
\bibitem[Thompson \& Madau(2000)]{tm00} Thompson, C., \& Madau, P.\ 2000, \apj, 538, 105 
\bibitem[Thompson et al.(2007)]{tmr07} Thompson, C., M{\'e}sz{\'a}ros, P., \& Rees, M.~J.\ 2007, \apj, 666, 1012 
\bibitem[Thompson \& Gill(2014)]{tg13} Thompson, C., \& Gill, R.\ 2014, ApJ, in press (arXiv:1310.2480)
\bibitem[Vishniac(1983)]{vishniac83} Vishniac, E.~T.\ 1983, \apj, 274, 152 
\bibitem[Waxman \& M{\'e}sz{\'a}ros(2003)]{waxman03} Waxman, E., \& M{\'e}sz{\'a}ros, P.\ 2003, \apj, 584, 390 
\bibitem[Zhang \& Yan(2011)]{zhang11} Zhang, B., \& Yan, H.\ 2011, \apj, 726, 90
\end{thebibliography}
\end{document}